\documentclass[useAMS]{mn2e}
\usepackage{epsfig}
\usepackage{amsmath}
\usepackage{rotating}
\newcommand{\be}{\begin{equation}}
\newcommand{\beq}{\begin{equation}}
\newcommand{\ba}{\begin{eqnarray}}
\newcommand{\ee}{\end{equation}}
\newcommand{\eeq}{\end{equation}}
\newcommand{\ea}{\end{eqnarray}}
\newcommand{\msun}{\rm M_{\odot}\hspace{1mm}}

\def\lsim{\rlap{$<$}{\lower 1.0ex\hbox{$\sim$}}}

\def\gsim{\rlap{$>$}{\lower 1.0ex\hbox{$\sim$}}}

\title[WL and GW Standard Sirens]{Cosmology with Standard Sirens: the Importance of \\ the Shape of the Lensing Magnification Distribution}
\author[C. Shang and Z. Haiman]{Cien Shang$^{1}$\thanks{E-mail: cien@phys.columbia.edu, zoltan@astro.columbia.edu} and Zolt\'an Haiman$^{2}$\\
$^1$Department of Physics, Columbia University, 550 West 120th Street, New York, NY 10027, USA; cien@phys.columbia.edu\\ 
$^2$Department of Astronomy, Columbia University, 550 West 120th Street, New York, NY 10027, USA; zoltan@astro.columbia.edu\\
}

\begin{document}

\date{\today}
\pagerange{\pageref{firstpage}--\pageref{lastpage}} \pubyear{2009}

\maketitle

\label{firstpage}

\begin{abstract}
The gravitational waves (GWs) emitted by inspiraling binary black
holes, expected to be detected by the {\it Laser Interferometer Space
Antenna (LISA)}, could be used to determine the luminosity distance to
these sources with the unprecedented precision of $\lsim 1$\%.  We
study cosmological parameter constraints from such standard sirens, in
the presence of gravitational lensing by large--scale structure.
Lensing introduces magnification with a probability distribution
function (PDF) whose shape is highly skewed and depends on
cosmological parameters.  We use Monte-Carlo simulations to generate
mock samples of standard sirens, including a small intrinsic scatter,
as well as the additional, larger scatter from lensing, in their
inferred distances. We derive constraints on cosmological parameters,
by simultaneously fitting the mean and the distribution of the
residuals on the distance {\it vs} redshift ($d_L-z$) Hubble diagram.
We find that for standard sirens at redshift $z\approx1$, the
sensitivity to a single cosmological parameter, such as the matter
density $\Omega_{m}$, or the dark energy equation of state $w$, is
$\sim 50\%-80\%$ tighter when the skewed lensing PDF is used, compared
to the sensitivity derived from a Gaussian PDF with the same
variance. When these two parameters are constrained simultaneously,
the skewness yields a further enhanced improvement (by $\sim 120\%$),
owing to the correlation between the parameters.  The sensitivity to
the amplitude of the matter power spectrum, $\sigma_8$ from the
cosmological dependence of the PDF alone, however, is $\sim 20\%$
worse than that from the Gaussian PDF.  The improvements for
$\Omega_m$ and $w$ arise purely from the non-Gaussian shape of the
lensing PDF; the dependence of the PDF on these parameters does not
improve constraints relative to those available from the mean $d_L-z$
relation.  At higher redshifts, the PDF resembles a Gaussian more
closely, and the effects of the skewness become less prominent.
These results highlight the importance of obtaining an accurate and
reliable PDF of the lensing convergence, in order to realize the full
potential of standard sirens as cosmological probes.
\end{abstract}

\section{Introduction}
\label{sec:introduction}

The proposed space-based GW detector {\it LISA}, sensitive to
frequencies between $\sim 10^{-5}$ to $\sim 0.1$ Hz, will be able to
detect massive black hole binary (MBHB) mergers out to redshifts
$z\gsim 5$.  In addition to providing information on black hole
physics and general relativity, observations of GWs by {\it LISA}
could be used as a probe of cosmology.  As pointed out in a pioneering
paper by Schutz (1986), GW observations of a binary system could yield
an accurate estimate of the luminosity distance to the source,
independent of assumptions about the masses and orbital parameters of
the binary members.  Recent analyses in the context of {\it LISA} show
that for many binaries, the luminosity distance could be measured to
percent--level precision (see Arun et al. 2009b for a review and
references).

Although a typical {\it LISA} source will be relatively poorly
localized on the sky (to $\sim 0.1$deg$^2$), when the spatial
information along the line of sight is taken into account, the number
of galaxies in the three-dimensional error volume will be reduced
significantly (Holz \& Hughes 2005; Kocsis et al. 2006).  Combining
this with possible tell-tale time-variable signatures (see Haiman et
al. 2009 for a review of several possibilities), it may become
feasible to identify an electromagnetic (EM) counterpart.  By
measuring the redshift of the counterpart, a gravitational version of
the Hubble diagram could be constructed.  This Hubble diagram, with a
relatively small intrinsic scatter, and spanning a large range in
redshift, can possibly impose tight constraints on cosmological
parameters (see, e.g., Arun et al. 2009a for a recent review and
references).

Unfortunately, however, gravitational lensing significantly changes
this picture. Standard sirens, just as type Ia supernovae (SNe Ia),
are (de)magnified by inhomogeneities in the matter distribution in the
foreground, which introduces an uncertainty in the measured
distance-redshift relation by up to $\sim 10$\% for high--redshift
($z\gsim 2$) events (e.g. Holz \& Hughes 2005; Kocsis et al. 2006).
In the case of SNe Ia, proposed missions, such as the
Supernova/Acceleration Probe (SNAP), are expected to find a few
thousand useful sources. The random lensing magnification errors then
average out, and even if they are unaccounted for, they have a
relatively modest ($\lsim1\%$) impact on cosmological
parameter-estimation, which can be ignored (Holz \& Linder 2005;
Sarkar et al. 2008).  The same strategy is unlikely for standard
sirens: although the expected {\it LISA} MBHB event rate is highly
uncertain, most models predict that it is significantly below the SNe
Ia rate, with perhaps tens of detections per year (e.g. Menou et
al. 2001; Sesana et al. 2007; Lippai et al. 2009; Arun et
al. 2009b). Furthermore, the EM counterpart may be identifiable for
only a fraction of these events.\footnote{The {\it Big Bang Observer
(BBO)}, a concept for a space mission to succeed {\it LISA}, could
detect a more than sufficient number of compact stellar binaries; this
would even allow a useful measurement of the spatial power spectrum of
the lensing convergence, which can provide additional cosmological
constraints (Cutler \& Holz 2009; see also Cooray et al. 2006 for the
same idea with Type Ia SNe).  In this paper, we will not consider the
possibility of such a large number ($\gsim 10^3$) of detectable
events.}

Motivated by the tremendous potential, in the absence of lensing, of
standard sirens for cosmology, there have been proposals to correct
for the effects of lensing of individual sources on a case-by-case
basis.  These proposals include measuring or constraining the
magnification using either photometric and spectroscopic properties of
foreground galaxies (e.g. J\"onsson et al. 2007), or the combination
of arcminute--scale shear and flexion maps (Shapiro et al. 2009; the
earlier work of Dalal et al. 2003, which only considered the shear,
concluded that only modest, $\sim 20\%$, corrections were feasible).
Both of these methods could reduce the lensing--induced distance
errors, in idealized cases, by a factor of up to $\sim$two.

Alternatively, several authors have investigated the possibility of
using lensing as a signal, rather than as noise, in probing cosmology
(Dodelson \& Vallinotto 2006, hereafter DV06; Linder 2008; see also
Wang et al. 2009). In particular, the variance of the lensing
magnification probability distribution function (PDF) depends on the
amplitude of density fluctuations and on the growth function, and
therefore could be used to constrain cosmological parameters such as
$\sigma_8$ and $\Omega_m$.  More specifically, DV06 showed that
$\sigma_8$ could be constrained to an accuracy of $\approx 5\%$ by
observations of 2000 SNe Ia.  The overall shape of the lensing
convergence distribution contains additional cosmological information,
beyond the variance (Wang et al. 2009). For example, Linder (2008)
emphasized that the theoretically possible minimum (de)magnification,
which occurs along an empty beam, depends on cosmology. Several
authors have indeed proposed fitting formulae for the lensing PDF,
derived from numerical simulations, which are self-similar, and depend
on cosmology only through the variance and the minimum magnification
(see below).

In this paper, we study the cosmological parameter constraints from
standard sirens, in the presence of lensing magnification.  In
general, lensing causes significant degradation of the constraints,
which includes increased uncertainties, as well as a possible bias, in
the parameters inferred from a given finite source sample.  If the
lensing PDF was Gaussian, and did not depend on the cosmological
parameters, this degradation could be estimated simply by the increase
in the variance of the inferred distance error to individual sources.
However, the lensing PDF is highly skewed, and it does depend on the
cosmological parameters.  Our focus here is to quantify the extent to
which these two features affect (and hopefully, mitigate) the
degradation of the constraints expected from {\it LISA}.

In the context of SNe, these questions have already been addressed in
detail by several authors, including the impact of lensing on inferred
dark energy parameters (e.g. Holz \& Linder 2005; Sarkar et al. 2008)
and on the normalization of the matter power--spectrum, $\sigma_8$
(e.g. DV06).  Here we consider, instead, a relatively small sample
($\sim$tens) of sources, with otherwise very small ($\lsim 1\%$)
distance errors, as expected from {\it LISA} standard sirens.  The
conclusions are not necessarily the same for such a standard siren
sample as for the SNe Ia.  This is because the probability
distribution of the inferred distances, which is the convolution of
intrinsic and lensing probability distributions, is much more skewed
in the case of standard sirens, due to the dominance of the highly
skewed lensing distribution. Additionally, depending on the statistic
being used, having fewer events can increase the impact of
non-Gaussianity on the inferred parameters.

A few works have also studied the cosmological utility of standard
sirens, and have included the effect of lensing, noting the
significant degradation they cause in the constraints (e.g. Holz \&
Hughes 2005; Dalal et al. 2006; Linder 2008).  Our present study adds
to these existing papers in the following ways: (i) in our analysis,
we include the dependence of the convergence PDF on cosmological
parameters, thus treating lensing as a potential source of signal,
rather than as pure noise; (ii) we include $\sigma_8$ among our
parameters, since the lensing PDF is particularly sensitive to this
parameter; (iii) we perform multiple Monte Carlo realizations of mock
standard-siren samples, incorporating the non-Gaussian shape of the
PDF, to obtain accurate estimates of the confidence intervals on the
inferred parameters; and (iv) we use more recent fitting formulae for
the lensing PDF, which improve the fit to numerical simulations.

The remainder of this paper is organized as follows.
In \S~\ref{sec:physics}, we briefly summarize the basic background
material required for our study, including information on both
standard siren distance measurements and on gravitational lensing.
\S~\ref{sec:mc} outlines the details of our Monte-Carlo simulation
procedure.
In \S~\ref{sec:results}, we present and discuss our main results, and
contrast these with the analogous results in the case of SNe Ia.
Finally, we summarize our main conclusions in \S~\ref{sec:conclusion}.
Throughout this paper, we adopt standard ${\rm \Lambda CDM}$ as our
fiducial cosmological model, with parameter values consistent with the
{\it WMAP} fifth year results (Komatsu et al. 2009), i.e.,
$\{\Omega_m, \Omega_{\Lambda},\Omega_b, h, \sigma_8, n_s\}=\{0.279,
0.721, 0.0462, 0.701, 0.817, 0.96\}$.  These values are in general
agreement with most recent, seventh year results, as well (Komatsu et
al. 2010).

\section{Standard siren distance measurement and gravitational lensing}
\label{sec:physics}

We follow the convention in the astronomical literature, and express
the luminosity distance $d_L(z)$, inferred from an observation of a
standard siren at redshift $z$, by the distance modulus,\footnote{We
note that in some applications, using fluxes, rather than distance
moduli, is preferable, since lensing acts linearly on the flux, and
therefore it does not shift the mean of a flux PDF (whereas lensing
can shift the mean of the magnitude PDF). However, as explained below,
in our analysis, we fit the entire PDF, and the two prescriptions
would lead to the same result.}
\begin{eqnarray}
\label{eqn:mtot}
m(z)=m_0(z)+\delta m_{\rm int}(z)+\delta m_{\rm len}(z).
\end{eqnarray}
Here $m_0(z)$ is the true distance modulus in a homogeneous universe,
depending only on the usual luminosity distance $d_L(z)$,
\begin{eqnarray}
\label{eqn:m0}
m_0(z)=5{\rm log_{10}}\left[\frac{d_L(z)}{10~{\rm pc}}\right],
\end{eqnarray}
and in a flat universe as assumed throughout this study,
\begin{eqnarray}
\label{eqn:dl}
d_L(z)=(1+z)c\int_{0}^{z}\frac{dz'}{H(z')},
\end{eqnarray}
where $c$ is the speed of light, and $H(z)$ is the Hubble parameter.

The other two terms in equation~(\ref{eqn:mtot}) vary from source to
source, and account for the intrinsic dispersion due to {\it LISA}'s
instrumental and background confusion noise ($\delta m_{\rm int}$),
and for the scatter caused by lensing magnification ($\delta m_{\rm
len}$).  In equation~(\ref{eqn:m0}), an arbitrary calibration constant
has been omitted. We furthermore ignore contributions to the scatter
from peculiar velocities, which are only important at low redshifts
($z\lsim 0.3$) and also neglect any other source of systematic errors,
which, of course, need to be included in the actual data analysis.

For a standard siren, the intrinsic error $\delta m_{\rm int}(z)$ is
usually very small, but the distribution of $\delta m_{\rm int}(z)$
depends on properties of the detector and of the MBHB in a
non--trivial manner. In the most general case, the width of the
intrinsic error distribution is a function of 17 parameters (Vecchio
2004; Holz \& Hughes 2005; Kocsis et al. 2006; Trias and Sintes 2008).
Owing to the complexity of the problem and to computational
limitations, the distribution of intrinsic luminosity distance errors
has been estimated for only a few discrete choices of redshifts and BH
masses. To scale the intrinsic error to an arbitrary redshift and mass
from a base value, we adopt the same strategy as in Kocsis et
al. (2006),
\begin{eqnarray}
\label{eqn:dlerror}
\frac{\delta d_L(M, z)}{d_L(M,
z)}=\left[\frac{(S/N)(M,z)}{(S/N)(M_0,z_0)}\right]^{-1}\frac{\delta
d_L(M_0, z_0)}{d_L(M_0, z_0)},
\end{eqnarray}
where $S/N$ is the expected signal-to-noise ratio (SNR) for the
detection of GWs. Equation (\ref{eqn:dlerror}) is motivated by the
fact that the signal amplitude is inversely proportional to the
luminosity distance. In computing the SNR, we have assumed one year
observation time prior to coalescing, and used the same noise spectrum
as in Lang \& Hughes (2006), which includes instrumental
noise\footnote{Generated by the on-line sensitivity curve generator,
http://www.srl.caltech.edu/~shane/sensitivity.} and confusion noise
(from both extragalactic and galactic sources).  Limited by the
low-frequency noise wall, the visible time of very massive MBHBs can
be less than one year, rendering equation~(\ref{eqn:dlerror})
inaccurate at large BH mass $M$. Nevertheless, this, as well as other
details of the intrinsic error distribution, make very little
difference to the results of the present study, as the intrinsic error
is smaller than the lensing error.  The base value of the luminosity
distance error is chosen to be $2.88\times 10^{-3}$ for an equal-mass
binary ($M_1=M_2=3\times 10^5 {\rm M_{\odot}}$) at $z=1$, computed by
Klein et al. (2009) using the full second post-Newtonian (2PN)
gravitational waveform, and also including spin-orbit precession.  The
variance of $\delta m_{\rm int}$, $\sigma^2_{\rm int}$, is related to
the fractional distance error $\delta d_L/d_L$ through
equation~(\ref{eqn:dl}),
\begin{eqnarray}
\label{eqn:sint}
\sigma_{\rm int}=5\delta {\rm log_{10}}d_L \approx 2.17\frac{\delta
d_l}{d_L}.
\end{eqnarray}

The lensing error $\delta m_{\rm len}(z)$ is directly related to the
lensing magnification $\mu$,
\begin{eqnarray}
\label{eqn:mlen}
\delta m_{\rm len}=-2.5{\rm log_{10}}\mu=5{\rm log_{10}}(1-\kappa),
\end{eqnarray}
where $\kappa$ is the convergence.  The PDF of $\kappa$, $P(\kappa)$,
is known to be skewed, since there exists a minimum convergence,
$\kappa_{\rm min}$, corresponding to an empty path between the source
and the observer.  Since in this paper, we are primarily interested in
quantifying the effects of the non-Gaussian tails of the PDF, we have
to rely on estimates of the PDF in cosmological simulations.  Using
the results of N-body simulations in White (2005), Wang et al. (2009)
compared three different models for $P(\kappa)$, proposed by Taruya et
al. (2002, i.e. the log-normal distribution), Wang et al. (2002,
hereafter W02) and Das \& Ostriker (2006, hereafter DO06). They
concluded that the fitting formulae proposed by DO06, $P_{\rm DO06}$
best fit the simulation data, while the formulae proposed by W02,
$P_{\rm W02}$, over-predict the tail of the convergence distribution.
In the following, we therefore adopt $P_{\rm DO06}$ as the ``true''
distribution,
\begin{eqnarray}
\label{eqn:ptrue}
&P_{\rm true}(\kappa)&\equiv P_{\rm DO06}(\kappa)=N\\\nonumber
&\times& {\rm exp}\left\{-\frac{[{\rm ln}(1+\kappa/|\kappa_{\rm
      min})+\Sigma^2/2|)]^2}{2\Sigma^2}\right.\\\nonumber
&\times&\left.[1+A/(1+\kappa/|\kappa_{\rm
    min}|)]\right\}\\\nonumber
&\times& \frac{d\kappa}{|\kappa_{\rm min}|+\kappa}.
\end{eqnarray}
Here, the three parameters $N$, $\Sigma$ and $A$ are fixed by three
constraint equations,
\begin{eqnarray}
\label{eqn:constraint1}
\int_{\kappa_{\rm min}}^{\infty}P_{\rm true}(\kappa)d\kappa=1,
\end{eqnarray}
\begin{eqnarray}
\label{eqn:constraint2}
\int_{\kappa_{\rm min}}^{\infty}\kappa P_{\rm true}(\kappa)d\kappa=0,
\end{eqnarray}
\begin{eqnarray}
\label{eqn:constraint3}
\int_{\kappa_{\rm min}}^{\infty}\kappa^2 P_{\rm
  true}(\kappa)d\kappa=\sigma_{\kappa}^2,
\end{eqnarray}
where $\sigma_{\kappa}^2$ is the variance of $\kappa$. Note, in
particular, that $P_{\rm DO06}$ depends on cosmology only through
$\kappa_{\rm min}$ and $\sigma^2_{\kappa}$. These, in turn, are
determined from the equations
\begin{eqnarray}
\label{eqn:kmin}
\kappa_{\rm
  min}=-\frac{3}{2}\Omega_m\left(\frac{H_0}{c}\right)^2\int_{0}^{\chi_s}d\chi
  (1+z)\frac{\chi(\chi_s-\chi)}{\chi_s},
\end{eqnarray}
and
\begin{eqnarray}
\label{eqn:kvar}
\sigma^2_{\kappa}&=&\frac{9\pi}{4}\left(\frac{\Omega_m
  H_0^2}{c^2}\right)^2\int_{0}^{\chi_s}d\chi
  \left[(1+z)\frac{\chi(\chi_s-\chi)}{\chi_s}\right]^2\\\nonumber &
  \times & \int_{0}^{\infty}\frac{dk}{k}\Delta^2(k,z),
\end{eqnarray}
where $\chi$ is the comoving distance to redshift $z$, $\chi_s$ is the
comoving distance to the source, and $\Delta^2(k,z)$ is the
dimensionless non--linear matter power spectrum, evaluated in this
study using the algorithm of Smith et al. (2003).  Similar to
equation~(\ref{eqn:sint}), the variance of $\delta m_{\rm len}$ could
then be computed from $\sigma_{\kappa}^2$. In the case of a Gaussian
distribution, using equation~(\ref{eqn:mlen}), we have
\begin{eqnarray}
\label{eqn:slen}
\sigma_{\rm len}\approx 2.17\sigma_{\kappa}.
\end{eqnarray}

\begin{figure}
\begin{tabular}{c}
\rotatebox{-0}{\resizebox{80mm}{!}{\includegraphics{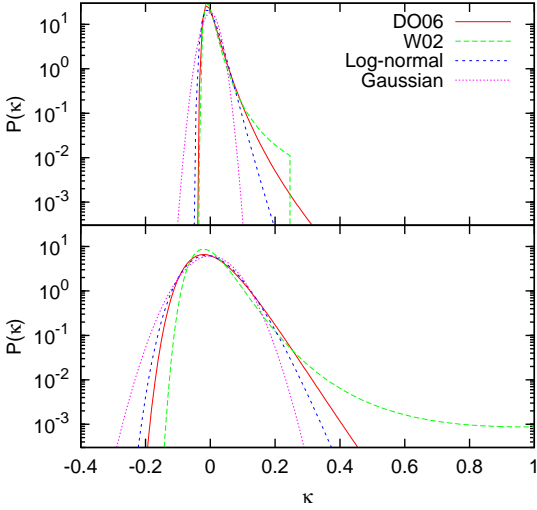}}}
\end{tabular}
\caption{PDFs of lensing convergence for sources at $z=1$ (upper
  panel) and $z=4$ (lower panel), from fitting formulae proposed by
  Das \& Ostriker (2006) and by Wang et al. (2002), compared to a
  log-normal and a Gaussian distribution, as labeled. Note that
  $P_{\rm W02}$ is sharply truncated in order to satisfy the
  constraint equation (\ref{eqn:constraint3}); the cutoff of $z=1(4)$
  is at $\kappa=0.25$ (1.46, outside the $\kappa$ range of the
  figure). 
  The low--redshift PDFs are more skewed than the high redshift ones,
  and $P_{\rm W02}$ predicts a significantly larger high-$\kappa$ tail
  than $P_{\rm DO06}$.}
\label{fig:pk}
\end{figure}

\begin{figure}
\begin{tabular}{c}
\rotatebox{-0}{\resizebox{80mm}{!}{\includegraphics{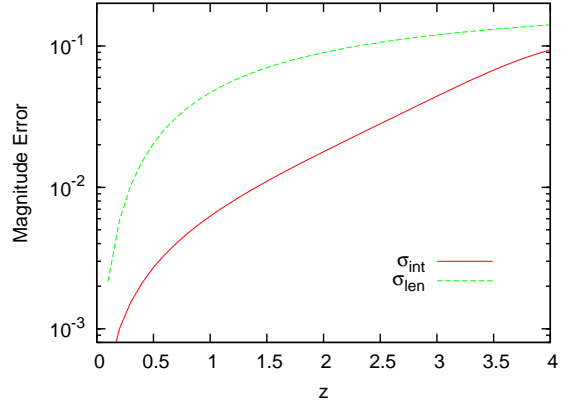}}}
\end{tabular}
\caption{The standard deviation of the intrinsic and lensing error
  distributions. The latter assumes an equal--mass binary
  ($M_1=M_2=3\times 10^5 {\rm M_{\odot}}$).  In contrast to SNe Ia,
  the intrinsic error for a typical standard siren at redshift $z\lsim
  4$ is a factor of $\gsim$two smaller than its lensing error (and
  nearly an order of magnitude smaller at $z\sim 1$).}
\label{fig:dispersion}
\end{figure}

It is important to note here that, in comparing fits to simulation
data, Wang et al. (2009) treated $\kappa_{\rm min}$ as a free
parameter. When $\kappa_{\rm min}$ is fixed at its theoretical value
from equation~(\ref{eqn:kmin}), $P_{\rm DO06}$ was, in fact, found to
under-predict the high--$\kappa$ tail of the PDF. The true PDF then
lies in--between $P_{\rm DO06}$ and $P_{\rm W02}$. Besides the
high--$\kappa$ tail, other issues with $P_{\rm W02}$ include
discontinuities and violating flux conservation at $z\lsim 0.6$
(Zentner \& Bhattacharya 2009).  We therefore adopt the fitting
formulae from DO6, despite their imperfect fit to the simulation data.
To illustrate the differences between the various PDFs, in
Figure~\ref{fig:pk}, we show $P_{\rm DO02}(\kappa)$, $P_{\rm
W02}(\kappa)$, a log-normal PDF, and a Gaussian PDF for sources at
$z=1$ (upper panel) and $z=4$ (lower panel).  The cutoff of $P_{\rm
W02}(\kappa)$ at $\kappa=0.25$ is imposed in order to satisfy the
constraint equation (\ref{eqn:constraint3}); $P_{\rm W02}(\kappa)$ of
$z=4$ is also truncated but the truncation is outside the $\kappa$
range shown in the figure.  As Figure~\ref{fig:pk} shows, $P_{\rm
DO6}$ and $P_{\rm W02}$ are both skewed, and the skewness is more
pronounced at lower redshift.  In Figure~\ref{fig:dispersion}, we show
the r.m.s. magnitude of the intrinsic and lensing dispersions as a
function of redshift.  For the latter, we assumed an equal--mass
binary with $M_1=M_2=3\times 10^5 {\rm M_{\odot}}$, which is close to
the best case for {\it LISA}.  At $z\lsim 4$, the intrinsic dispersion
is always a factor of $\gsim$two smaller than the lensing dispersion.

In reality, there will most likely be a wide distribution of masses
and mass ratios among the BH binaries detected by {\it LISA}.  These
distributions are highly uncertain, and for simplicity, we avoid
modeling it in this paper.  When both masses are in the range of $\sim
10^5 -10^6~\msun$, the variation of the intrinsic error with the mass
ratio is relatively modest (within a factor of $\sim 2$; Klein et
al. 2009), and the fiducial accuracies we chose are typical of these
binaries.  For binaries with component masses significantly outside
this optimal range, the intrinsic error will be larger than assumed
here, and for some events, it will therefore likely become comparable
(or larger) than the lensing error.  Nevertheless, for most events, we
expect the lensing error to remain dominant, at least at redshifts as
low as $z\sim 1$, where the differences shown in
Figure~\ref{fig:dispersion} are nearly an order of magnitude.
\begin{figure}
\begin{tabular}{c}
\rotatebox{-0}{\resizebox{80mm}{!}{\includegraphics{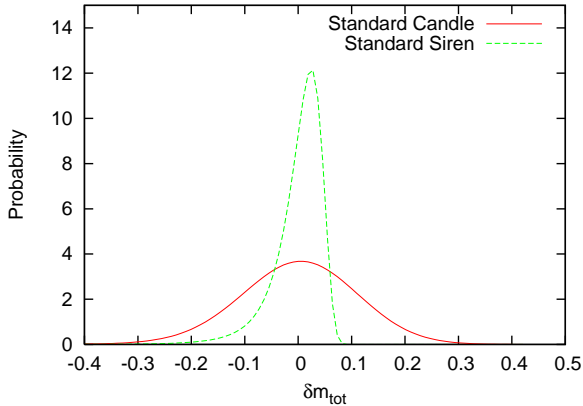}}}
\end{tabular}
\caption{Probability distribution of the total magnitude error of
  standard candles (solid [red] curve) and of standard sirens (dashed
  [green] curve) at $z=1$. The intrinsic errors of both standard
  candles and of standard sirens are assumed to have Gaussian
  distributions (in magnitude), with a standard deviations of 0.1 mag,
  and $2.88\times 10^{-3}$ mag, respectively; this Gaussian is
  convolved with the DO06 lensing PDF (shown separately in
  Figure~\ref{fig:pk}).}
\label{fig:pm}
\end{figure}

In Figure~\ref{fig:pm}, we show the overall probability distribution
of the magnitude error, given by the convolution of the distributions
of $\delta m_{\rm int}$ and $\delta m_{\rm len}$.  The solid (red)
curve is for standard candles, assuming an intrinsic dispersion that
is Gaussian (in magnitude) with a standard deviation of 0.1 mag (a
canonical value adopted by DV06 and in many other papers; e.g. Holz \&
Linder 2005; Zentner et al. 2009).  The dashed (green) curve is for
standard sirens, again assuming a Gaussian intrinsic error
distribution with a standard deviation of $2.88\times 10^{-3}$ mag.
In both cases, the source is assumed to be at redshift $z=1$.  As the
figure shows, the overall error distribution of standard sirens
remains much more skewed than that of standard candles.

We note that the intrinsic distribution of $\delta m_{\rm int}$ may
well be skewed, as well.  However, as long as $\delta m_{\rm int}$ is
several times smaller than $\delta m_{\rm len}$, the non-Gaussianity
in $\delta m_{\rm int}$ should not have a significant impact on our
conclusions.

\section{Likelihood analysis and Monte-Carlo simulation}
\label{sec:mc}

We next use Monte-Carlo (MC) simulations to derive confidence
intervals on the parameters inferred from mock samples of standard
sirens.  We follow a procedure similar to DV06, consisting of the
following steps:

(1) First, we generate random redshifts for $N_{\rm event}$ standard
sirens, distributed uniformly in a bin of width $\Delta z = 1$,
centered at redshift $z_{\rm cent}$. The number of events, $N_{\rm
  event}$, in each realization is varied, ranging from 40 to 2000.
The lower end of this range corresponds approximately to a
(pessimistic) expectation for the total number of {\it LISA} events,
while the upper end corresponds to the number of SNe expected in
future SN surveys such as SNAP/JDEM (allowing us to contrast standards
sirens to SNe).  The redshift $z_{\rm cent}$ is set at 1, 2, 3, and 4,
i.e., the events are between redshift 0.5 to 4.5. At $z\lsim0.5$, very
few standard sirens are expected, and the errors from peculiar
velocities become comparable to those from lensing (e.g. Kocsis et
al. 2006) and would have to be included in the analysis.  At
$z\gsim4.5$, the lensing PDF approaches a Gaussian shape, and the
effects of non-Gaussianity become negligible, as we will see in
\S~\ref{sec:results}.

(2) For each event $i$, we compute the true distance modulus using
equation (\ref{eqn:m0}).  We then draw two random values of $\kappa$,
one from the distribution of $P_{\rm true}(\kappa)$ and the other from
$P_{\rm Gass}(\kappa)$ whose variance is set to be the same as
predicted in equation~(\ref{eqn:constraint3}), and convert these
$\kappa$ values to $\delta m_{\rm len}$.  We also draw a random
$\delta m_{\rm int}$ from a Gaussian distribution with zero mean and
variance $\sigma_{\rm int}$.

(3) We add $\delta m_{\rm int}$ and $\delta m_{\rm len}$ to $m_0$ to
form two sets of mock data, ($z_i$, $m_i$), corresponding to the two
choices of the convergence PDF.

(4) We then perform two separate likelihood analyses on the mock data,
assuming the convergence PDF, either true or Gaussian, is known.  The
distribution of $\delta m_{\rm int}$ is then convolved with that of
$\delta m_{\rm len}$ to obtain the PDF of the total error, $P_{\rm
  tot}$.  In this analysis, in difference from DV06, we assume that
$\sigma_{\rm int}$ is known and is fixed at the ``correct'' value
(i.e. the same value used in step (2) above to generate the mock
sample).  As mentioned above, in our case, the intrinsic dispersion is
sub-dominant compared to the lensing dispersion, and we therefore do
not expect marginalizing over this intrinsic dispersion will modify
our results.

(5) In each analysis, we vary either one or two cosmological
parameters, and find the parameter value(s), $p^{\rm max}$, that
maximizes the likelihood $\cal{L}$, or equivalently minimizes the
quantity $\Xi\equiv -2{\rm ln}\cal{L}$. For a normalized $P_{\rm
  tot}$,
\begin{eqnarray}
\label{eqn:2lnl}
\Xi\equiv -2{\rm ln}{\cal{L}}=-2\sum_{i}{\rm ln}P_{\rm tot}(m_i-m_0(z_i)),
\end{eqnarray}
where terms independent of cosmology have been omitted (Marshall et
al. 1983).  In the limit of Gaussian probability distribution $P_{\rm
tot}$, $-2{\rm ln}\cal{L}$ becomes similar to the usual $\chi^2$
statistic.

6) We repeat the steps (1)-(5) 1000 times to produce two distributions
of $p^{\rm max}$, corresponding to the two distributions of $\kappa$.

7) Finally, we evaluate the effects of the convergence PDF on the
cosmological constraints by comparing these two distributions of
$p^{\rm max}$.

The above approach allows us to quantify the impact of the
non-Gaussian shape of the convergence PDF on the distribution of the
parameter estimates.  It is useful to contrast here the above approach
to some others that are often used in the literature to forecast
parameter errors.  The Fisher-matrix formalism (Tegmark et al. 1997)
is commonly adopted when the number of parameters to be estimated
simultaneously is too large.  This method assumes that the likelihood
function is Gaussian, both as a function of the parameters to be
estimated and as a function of the observables. In the present case,
both of these assumptions are false (our results will be explicitly
compared to those from a Fisher matrix below).  Another common
practice is to compute the likelihood, as a function of the
parameters, around a single realization of the mock data -- usually,
either the most likely realization (see, e.g. Figure 6 in Holz \&
Hughes 2005), or in a ``typical'' realization (see, e.g., DV06).  This
avoids the assumption that the likelihood is Gaussian in the
parameters, but does not fully capture the non-Gaussian shape of the
distribution as a function of the observables.

In the context of SNe, Holz \& Linder (2005) have used full Monte
Carlo simulations to quantify the impact of lensing on the
distribution of the inferred best--fit parameters, similar to our
procedure above.  Because of the large intrinsic dispersion of the
SNe, lensing is less important to begin with, and the effect of the
non-Gaussianity is also relatively modest (see their Figures 8 and 9).
As far as we are aware, the same approach of using full Monte Carlo
simulations have not been used to study the impact of lensing for
standard sirens, for which the intrinsic dispersion is smaller, and
for which the effects of lensing, as well as of non-Gaussianities, are
more important.

\section{Results and Discussion}
\label{sec:results}

\subsection{Sensitivity to a single cosmological parameter}
\label{subsec:1p}

We begin by investigating the sensitivity of standard sirens to a
single parameter. We focus on the degradation caused by lensing, and
how this degradation is affected (hopefully, mitigated) by the
non-Gaussian shape of the lensing PDF and by its dependence on
cosmological parameters.

\begin{figure*}
\begin{tabular}{c}
\rotatebox{-0}{\resizebox{170mm}{!}{\includegraphics{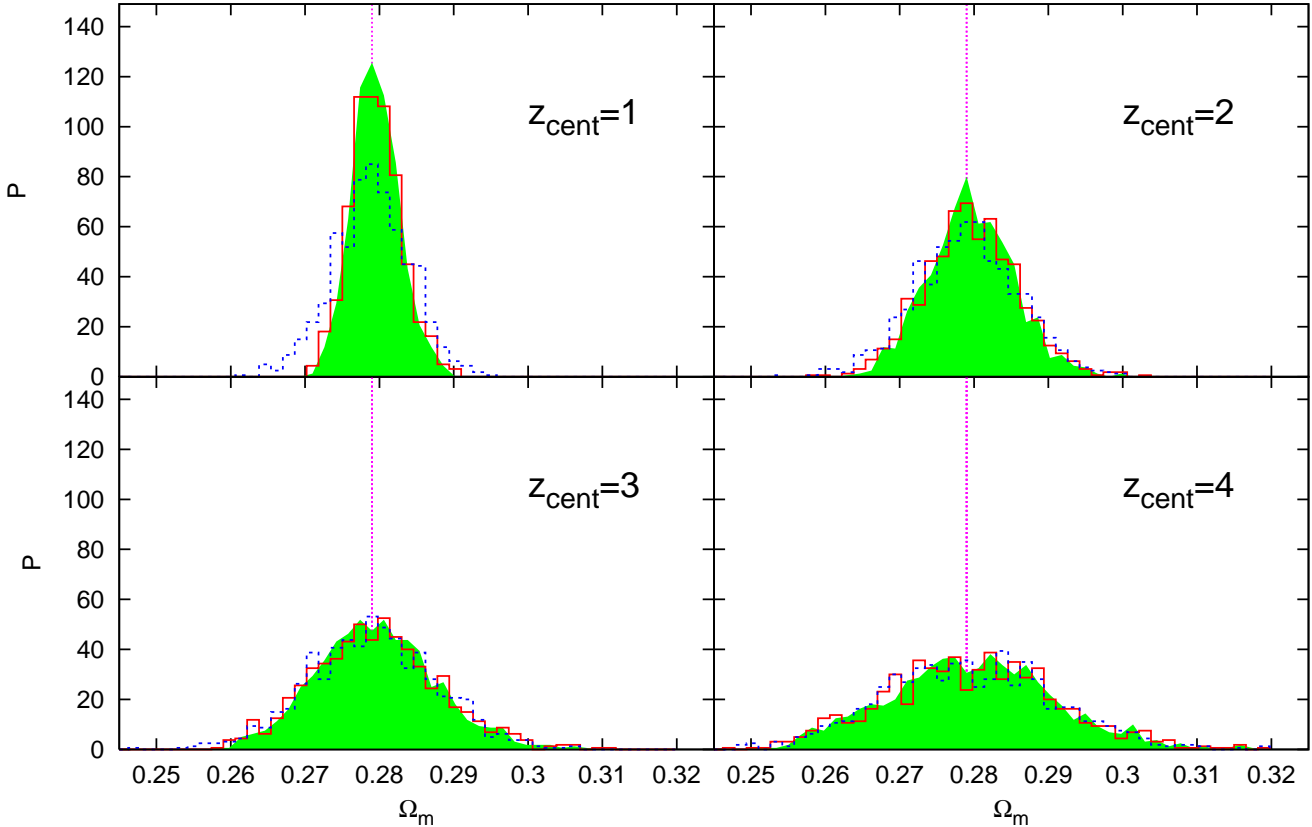}}}
\end{tabular}
\caption{Distributions of $\Omega_m$ that maximize the likelihood. The
  solid [red] curve, and the dashed [blue] curves are the
  distributions when $P_{\rm true}(\kappa)$ and $P_{\rm Gass}(\kappa)$
  are used, respectively. The shaded [green] region also shows the
  distribution for $P_{\rm true}(\kappa)$, but with the PDF fixed
  using the fiducial values of the cosmological parameters. The
  vertical dotted lines in this and following similar figures indicate
  the fiducial value of the cosmological parameter being
  constrained. The number of events in each realization is 40.}
\label{fig:om1}
\end{figure*}

\begin{figure*}
\begin{tabular}{c}
\rotatebox{-0}{\resizebox{170mm}{!}{\includegraphics{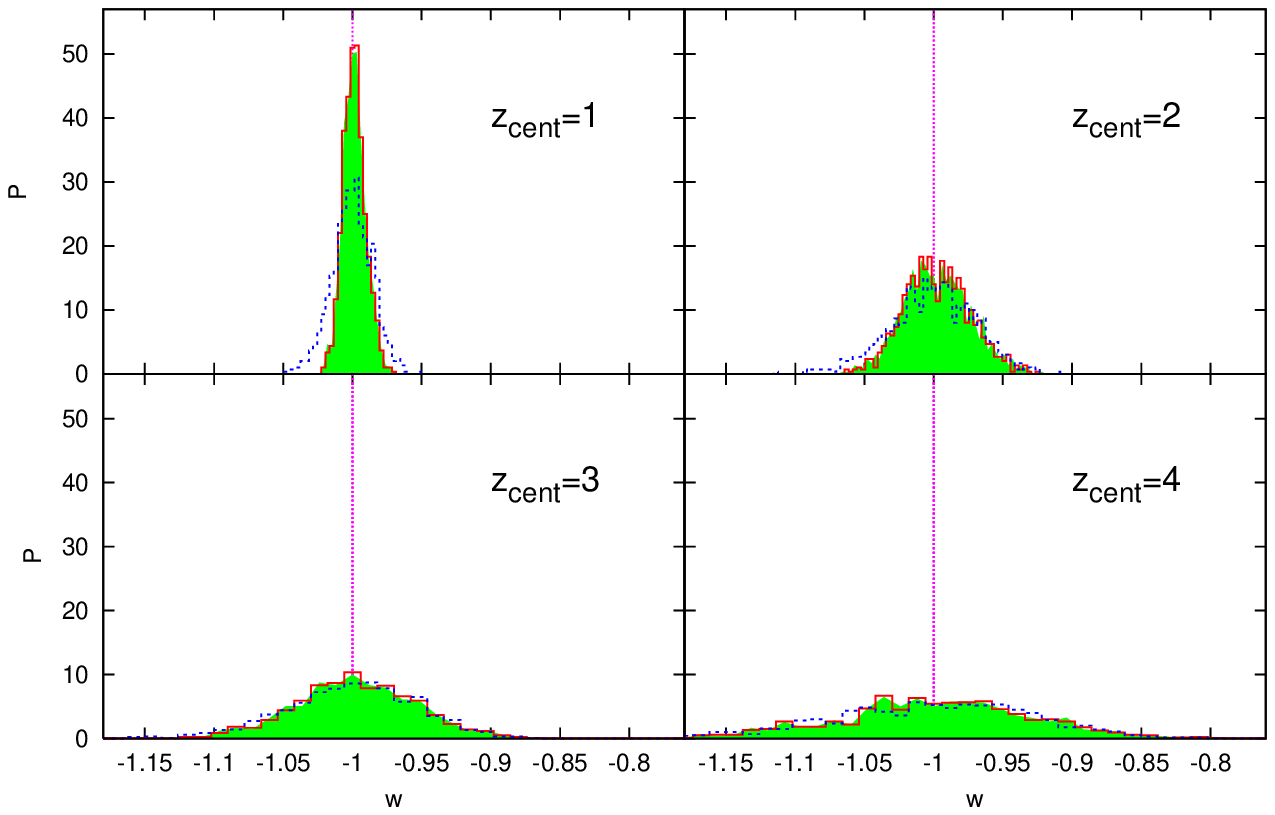}}}
\end{tabular}
\caption{Distributions of best--fit $w$ values when the true (solid
  [red] curve) and Gaussian (dashed [blue] curve) PDF are used in the
  likelihood analysis. 1000 mock samples are generated with 40 events
  in each realization, distributed randomly in redshift within a bin
  centered at $z_{\rm cent}=1$, 2, 3 or 4, as labeled.  As in
  Figure~\ref{fig:om1}, the shaded [green] region shows results when
  the lensing PDF artificially uses the fixed values of the fiducial
  cosmological parameters.  }
\label{fig:w1}
\end{figure*}

\subsubsection{Matter density $\Omega_m$}
\label{subsec:varyomega}

In the first example, we study the constraints on $\Omega_m$ while
other parameters are held fixed at their fiducial values. The number
of events in each realization is 40, and $z_{\rm cent}$ is set either
at 1, 2, 3, or 4. The distributions of best--fit $\Omega_m$ values are
shown in Figure~\ref{fig:om1}, where the solid (red) curves and dashed
(blue) curves correspond to the results when $P_{\rm true}(\kappa)$
and $P_{\rm Gass}(\kappa)$ are used in the likelihood analysis,
respectively. The shaded (green) region also shows the distribution
for $P_{\rm true}(\kappa)$, but with the PDF fixed using the fiducial
values of the cosmological parameters. The true (fiducial) $\Omega_m$
is indicated by the vertical dotted lines.

The parameters of these distributions are summarized quantitatively in
Table (\ref{tbl:om1}), which lists the peak, 68\% and 95\% errors of
best--fit $\Omega_m$ and the ratios of the errors resulting from the
two different assumptions about the convergence PDF (Gaussian
vs. non-Gaussian).  The second to last column shows the degradation
due to the lensing, indicated by the ratio of 68\% errors with and
without lensing.  For a comparison of constraining power of individual
standard sirens and that of SNe Ia, the last column of Table
(\ref{tbl:om1}) shows how many standard sirens need to be observed in
each redshift bin in order to obtain the same constraints on
$\Omega_m$ as two thousand SNe Ia at redshifts between 0.5 and
1.5. Within the parentheses are the corresponding numbers of events
but in the hypothetical case that lensing is either absent or can be
perfectly corrected for.  In particular, an event in the first
redshift bin on average has cosmological constraining power similar to
$2000/114\approx 18$ SNe Ia.

Several interesting conclusions can be drawn from Figure~\ref{fig:om1}
and Table~\ref{tbl:om1}.  First, consistent with previous works, our
results show that lensing severely degrades the constraints from
standard sirens, and that this degradation is worse than for SNe Ia,
due to the small intrinsic dispersion of standard sirens. At $z\sim1$,
lensing increases the 68\% error by a factor of 4.71 for standard
sirens, while only by 25\% for SNe Ia.  We also note that the
best--fit values of $\Omega_m$ are essentially unbiased; this is
simply because we have assumed that the shape of the lensing PDF is
known ab--initio (we will relax this assumption in
\S~\ref{subsec:bias} below).

Second, the constraints on $\Omega_m$ become worse as $z_{\rm cent}$
increases. This is because both the intrinsic and the lensing
dispersions rapidly increase with redshift, as shown in Figure
(\ref{fig:dispersion}). When the number of events in a realization is
fixed, the error on $\Omega_m$ is roughly proportional to the total
dispersion of the measured distance modulus,
\begin{eqnarray}
\label{eqn:explain1}
\delta \Omega_m &\sim& \delta m \left|\frac{d m}{d\Omega_m}\right|^{-1} 
\propto \delta m \left|\frac {1}{d_L} \frac{d d_L}{d \Omega_m}\right|^{-1} \\
\nonumber &\sim& 2 \Omega_m \delta m.
\end{eqnarray}
In the above, we have used the approximation $H(z)\approx H_0
\sqrt{\Omega_m (1+z)^3}~(z\gsim 1)$, which allows us to move
$\Omega_m$ out of the integral of $d_L$, and cancel the redshift
dependence.  Taking the values of $\sigma_{\rm true}$ in Table
(\ref{tbl:om1}), and using the scaling that the error on a
cosmological parameter is inversely proportional to the square root of
the number of events, our results show that an event at $z=1$ has
roughly the same contribution in constraining $\Omega_m$ as $(1.14
\times 10^{-2}/3.49 \times 10^{-3})^2\approx 11$ events at $z=4$.

The third and most interesting conclusion is that the constraints in
the true PDF case are tighter than those in the Gaussian PDF case. The
difference is as large as $\sim50\%$ in the lowest redshift bin, where
the true PDF is the most skewed and is furthest from a Gaussian shape,
as clearly seen in Figure~\ref{fig:pk}. This illustrates the
importance of the shape of the convergence PDF, which, if not
appropriately treated, could result in a biased estimate of the
constraints. For example, the popular Fisher matrix technique, as we
have checked, gives constraints on $\Omega_m$ that are very close to
our Gaussian case. For any attempts of correcting for lensing on a
case-by-case basis, this means that not only the variance, but the
full shape of the PDF needs to be considered in the analysis, in order
to evaluate the gain from such corrections. In general, these results
are good news for the prospects of using standard sirens in cosmology,
as the degradation from lensing is somewhat less severe than estimated
from the variance alone.  We also note, however, that the benefits of
the non-Gaussianities largely disappear at redshifts beyond $z\gsim
2$, where the lensing PDF is closer to a Gaussian.

\subsubsection{Dark energy equation of state $w$}
\label{subsec:varyw}

We have also run MC simulations in which we varied the dark energy
equation of state parameter $w$, while holding all other parameters
fixed. The results are shown in Table (\ref{tbl:w1}) and
Figure~\ref{fig:w1}. The mock survey parameters are the same as in
Table~\ref{tbl:om1}. The results are qualitatively similar to those in
the case of $\Omega_m$. The benefits of the non-Gaussian shape are,
however, somewhat larger, with the degradation due to lensing smaller
by up to $\sim 80\%$ (in the lowest redshift bin) compared to the
Gaussian case. We also note that the errors on $w$ have a steeper
redshift dependence than those of $\Omega_m$. An event at $z=1$ has a
constraining power on $w$ equivalent to $(6.83 \times 10^ {-2}/8.08
\times 10^ {-3})^2 \approx 71$ events at $z=4$, compared to $\sim 11$
for $\Omega_m$. The transition from cosmic deceleration to
acceleration happens at $z\sim 1$; events at this redshift are
therefore especially sensitive probes of dark energy.

\begin{table*}
  \caption{The peak position ($\nu$) and the $68\%$ and $95\%$
    confidence levels for the distribution of the best-fit $\Omega_m$
    when the convergence PDF is assumed to be Gaussian (denoted by the
    subscript ``Gass'') and when it is assumed to follow the
    non--Gaussian DO06 distribution (denoted by the subscript
    ``true'').  The ratios of the 68\% errors in these two cases is
    also shown. Each realization generates a mock sample of 40
    standard sirens, distributed uniformly in redshift within a width
    $\Delta z= 1$, centered at $z_{\rm cent}=$ 1, 2, 3, or 4.  The
    second to last column shows the degradation caused by lensing by
    comparing the 68\% errors with and without lensing effects. The
    last column lists the number of standard sirens that would have
    similar constraining power (evaluated using the 68\% error) as
    2000 SNe Ia uniformly distributed between redshift 0.5 and
    1.5. Within the parentheses is the corresponding number of events
    in the absence of lensing or with the lensing effects perfectly
    corrected.}
    \label{tbl:om1}

\begin{center}
\begin{tabular}{c|c|c|c|c|c|c|c|c|c|c}
\hline \hline
 & \multicolumn{3}{c}{Gaussian
  PDF}&\multicolumn{3}{c}{True PDF}& & \\
$z_{\rm cent}$&$\nu_{\rm Gass}$&$\sigma_{68,\rm Gass}$&$\sigma_{95,\rm
 Gass}$&$\nu_{\rm true}$&$\sigma_{68,\rm true}$ &$\sigma_{95,\rm
 true}$ & $\frac{\sigma_{68,\rm Gass}}{\sigma_{68,\rm true}}$&
 $\frac{\sigma_{95,\rm Gass}}{\sigma_{95,\rm true}}$ & $\frac{\sigma_{68,\rm True}}{\sigma_{68,\rm nolens}}$ & $N_{\rm SN}$\\ 
\hline
1 & 0.279&$ 5.22 \times 10^ {-3}$&$ 1.06 \times 10^ {-2}$& 0.279&$ 3.49 \times 10^ {-3}$&$ 7.11 \times 10^ {-3}$& 1.50& 1.49& 4.71&$ 1.14 \times 10^ {2}( 8)$\\
2 & 0.279&$ 7.06 \times 10^ {-3}$&$ 1.42 \times 10^ {-2}$& 0.279&$ 6.24 \times 10^ {-3}$&$ 1.26 \times 10^ {-2}$& 1.13& 1.12& 4.57&$ 3.65 \times 10^ {2}( 27)$\\
3 & 0.279&$ 8.60 \times 10^ {-3}$&$ 1.73 \times 10^ {-2}$& 0.279&$ 8.52 \times 10^ {-3}$&$ 1.74 \times 10^ {-2}$& 1.01& 0.99& 3.00&$ 6.80 \times 10^ {2}( 118)$\\
4 & 0.285&$ 1.14 \times 10^ {-2}$&$ 2.27 \times 10^ {-2}$& 0.276&$ 1.14 \times 10^ {-2}$&$ 2.31 \times 10^ {-2}$& 1.00& 0.98& 1.83&$ 1.21 \times 10^ {3}( 565)$\\
\hline
\end{tabular}
\end{center}
\end{table*}

\begin{table*}{t}
  \caption{Peak, 68\% and 98\% errors of best--fit $w$ values, for different
    assumptions about the convergence PDF, as in Table~\ref{tbl:om1}. 
    The ratios of the errors for the different PDFs are also shown. Each
    realization includes 40 events.}
  \label{tbl:w1}
\begin{center}
\begin{tabular}{c|c|c|c|c|c|c|c|c}
\hline \hline
 & \multicolumn{3}{c}{Gaussian
  PDF}&\multicolumn{3}{c}{True PDF}& & \\
$z_{\rm cent}$&$\nu_{\rm Gass}$&$\sigma_{68,\rm Gass}$&$\sigma_{95,\rm
 Gass}$&$\nu_{\rm true}$&$\sigma_{68,\rm true}$ &$\sigma_{95,\rm
 true}$ & $\frac{\sigma_{68,\rm Gass}}{\sigma_{68,\rm true}}$&
 $\frac{\sigma_{95,\rm Gass}}{\sigma_{95,\rm true}}$\\ 
\hline
1 & -1.000&$ 1.43 \times 10^ {-2}$&$ 2.94 \times 10^ {-2}$& -1.000&$ 8.08 \times 10^ {-3}$&$ 1.66 \times 10^ {-2}$& 1.78& 1.77\\
2 & -0.994&$ 3.01 \times 10^ {-2}$&$ 6.06 \times 10^ {-2}$& -1.006&$ 2.33 \times 10^ {-2}$&$ 4.73 \times 10^ {-2}$& 1.29& 1.28\\
3 & -1.000&$ 4.76 \times 10^ {-2}$&$ 9.67 \times 10^ {-2}$& -1.000&$ 4.31 \times 10^ {-2}$&$ 8.85 \times 10^ {-2}$& 1.11& 1.09\\
4 & -0.964&$ 7.32 \times 10^ {-2}$&$ 1.47 \times 10^ {-1}$& -1.024&$ 6.83 \times 10^ {-2}$&$ 1.38 \times 10^ {-1}$& 1.07& 1.06\\
\hline
\end{tabular}
\end{center}
\end{table*}

\subsubsection{Power spectrum normalization $\sigma_8$}
\label{subsec:varysigma8}

\begin{figure*}
\begin{tabular}{c}
\rotatebox{-0}{\resizebox{170mm}{!}{\includegraphics{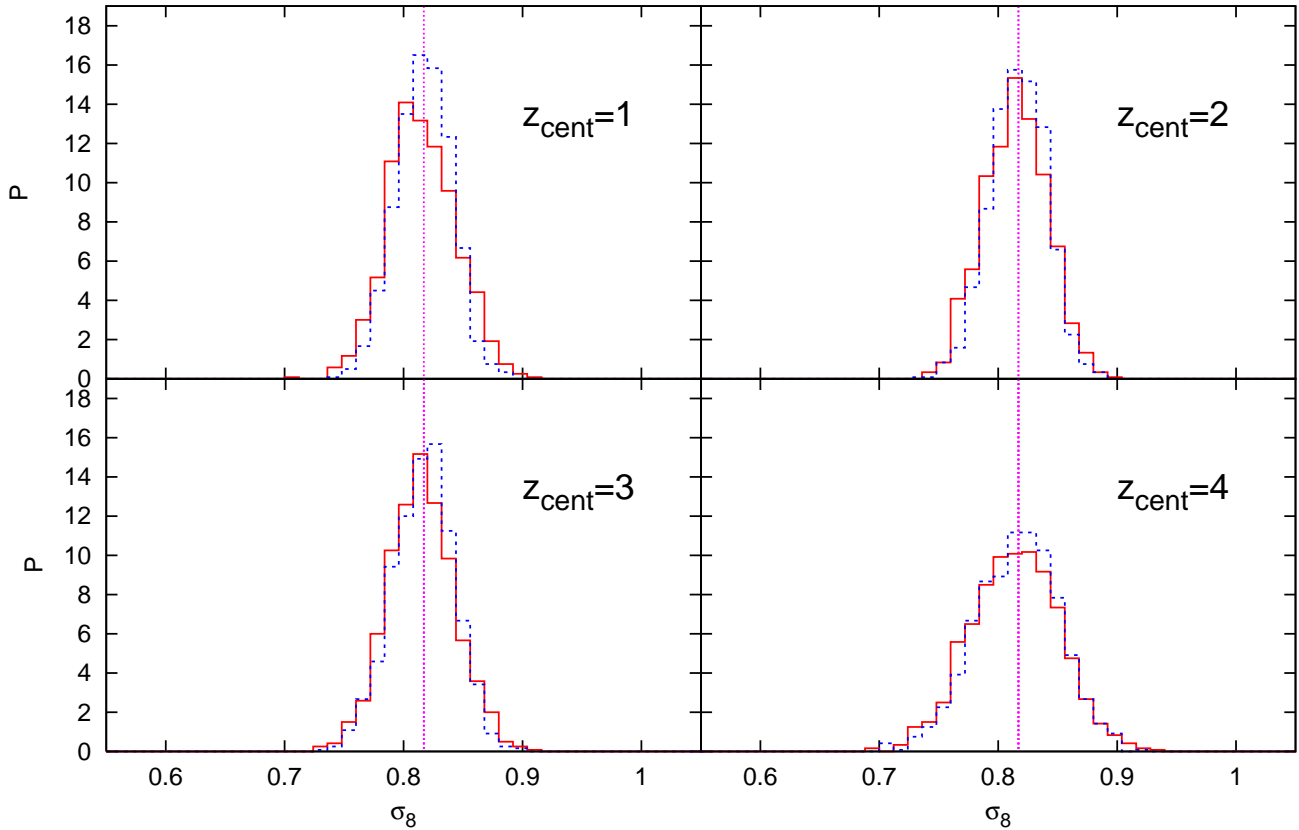}}}
\end{tabular}
\caption{Distributions of best--fit $\sigma_8$ values in 1000
  Monte--Carlo realizations when the true (solid [red] curve) and a
  Gaussian (dashed [blue] curve) PDF are used in the likelihood
  analysis. 400 events are generated in each realization, in a
  redshift bin centered at $z_{\rm cent}=1$, 2, 3, or 4, as labeled.}
\label{fig:s81}
\end{figure*}

We also studied $\sigma_8$, which is different from $\Omega_m$ and $w$
in that $\sigma_8$ has no effect on the mean distance modulus, and
therefore could not be as tightly constrained by {\it LISA} alone. To
obtain good constraints (for numerical convenience), four hundred
events are generated in each realization, and the results are shown in
Figure~\ref{fig:s81} and Table (\ref{tbl:s8}). Note that constraints
on $\sigma_8$, as opposed to those on $\Omega_m$ and $w$, have only a
very mild dependence on redshift: the 68\% errors are $\approx 0.025$
at $z=1, 2$ and 3 for the case of the Gaussian distribution. This
result could be understood by the following argument. First, we see
from equation (\ref{eqn:kvar}) that $\sigma^2_{\kappa}\propto
\Delta^2(k,z) \propto \sigma_8^2$, of which the latter proportionality
is only approximate because of nonlinear effects. This yields
\begin{eqnarray}
\label{eqn:explain2}
\frac{\delta\sigma_8^2}{\sigma_8^2}=\frac{\delta\sigma^2_{\kappa}}{\sigma^2_{\kappa}}
=\frac{(2/N_{\rm
    event})^{1/2}\sigma^2_{\kappa}}{\sigma^2_{\kappa}}=(2/N_{\rm
  event})^{1/2}\approx 0.071,
\end{eqnarray}
independent of redshift. When estimated from
equation~(\ref{eqn:explain2}) instead, $\delta \sigma_8\approx \delta
\sigma_8^2/2\approx 0.0236$, which is only slightly lower than the
values from the numerical calculation; the small difference can be
attributed to the intrinsic dispersion.

Another important difference is that non--Gaussianities {\it degrade},
rather than mitigate the lensing errors on $\sigma_8$, compared to the
Gaussian PDF case -- by $\sim 20\%$ in the lowest redshift bin. This
is in contrast to the situation for $\Omega_m$ and $w$, for which
non--Gaussianities tighten the constraints. The reason for this
finding will be investigated in \S~\ref{subsec:reason} below.

\begin{table*}
  \caption{Same as Table (\ref{tbl:w1}), except that $\sigma_8$ is
    being constrained and each realization produces 400 events.}
  \label{tbl:s8}
\begin{center}
\begin{tabular}{c|c|c|c|c|c|c|c|c}
\hline \hline
 & \multicolumn{3}{c}{Gaussian
  PDF}&\multicolumn{3}{c}{True PDF}& & \\
$z_{\rm cent}$&$\nu_{\rm Gass}$&$\sigma_{68,\rm Gass}$&$\sigma_{95,\rm
 Gass}$&$\nu_{\rm true}$&$\sigma_{68,\rm true}$ &$\sigma_{95,\rm
 true}$ & $\frac{\sigma_{68,\rm Gass}}{\sigma_{68,\rm true}}$&
 $\frac{\sigma_{95,\rm Gass}}{\sigma_{95,\rm true}}$\\ 
\hline
1 & 0.814&$ 2.42 \times 10^ {-2}$&$ 4.65 \times 10^ {-2}$& 0.814&$ 2.93 \times 10^ {-2}$&$ 5.76 \times 10^ {-2}$& 0.83& 0.81\\
2 & 0.814&$ 2.48 \times 10^ {-2}$&$ 4.76 \times 10^ {-2}$& 0.814&$ 2.78 \times 10^ {-2}$&$ 5.32 \times 10^ {-2}$& 0.89& 0.89\\
3 & 0.814&$ 2.64 \times 10^ {-2}$&$ 5.21 \times 10^ {-2}$& 0.814&$ 2.83 \times 10^ {-2}$&$ 5.74 \times 10^ {-2}$& 0.93& 0.91\\
4 & 0.826&$ 3.56 \times 10^ {-2}$&$ 7.01 \times 10^ {-2}$& 0.814&$ 3.78 \times 10^ {-2}$&$ 7.45 \times 10^ {-2}$& 0.94& 0.94\\
\hline
\end{tabular}
\end{center}
\end{table*}

\subsection{Information from the parameter-dependence of the PDF}
\label{subsec:parameterdependence}

As explained above, we have included two effects about the lensing PDF
in our analysis: (i) its non-Gaussian shape, and (ii) its dependence
on cosmological parameters.  It is important to clarify which of these
two effects was responsible for the improvement in the constraints on
$\Omega_m$ and $w$, relative to the case when the lensing PDF is
Gaussian and is considered pure noise (i.e. with no
parameter--dependence).  Assuming a Gaussian PDF and employing the
Fisher matrix method, Zentner \& Bhattacharya (2009), find that the
dispersion of SNe Ia moduli can help break degeneracies between dark
energy parameters.  However, the non-Gaussian, high-convergence tail
of the lensing PDF has been shown explicitly to contain cosmological
information, whose statistical accuracy in an all-sky map is
competitive with other cosmological probes (Wang et al. 2009).

We perform two tests in order to establish which of the two effects is
more important. First, we artificially fix the shape of the PDF with
the fiducial values of cosmological parameters throughout the
analysis. The results of this academic exercise are shown in
Figures~\ref{fig:om1} and \ref{fig:w1} by the shaded regions. This
brings only a marginal change in the distributions of the 
best--fit values compared to the fiducial case, showing that the
effects of the true lensing PDF come primarily through its
non-Gaussianity, rather than through its cosmology--dependence. 

This conclusion is confirmed by a second test, in which we
artificially fix the distance modulus, and re--derive constraints on
$\Omega_m$.  We find that the constraints in this case become by about
an order of magnitude worse. A closer investigation shows that for
$\Omega_m$, the cosmological dependence of the PDF, in fact, slightly
harms the constraints, while for $w$, it slightly improves the
constraints. The fact that the cosmological dependence of the PDF
harms the constraints on $\Omega_m$ must also be caused by the
non-Gaussian shape of the PDF. This is because for a Gaussian PDF, it
could be argued (e.g. using the Fisher matrix) that additional
information can only improve the sensitivity to a parameter (by
increasing the absolute value of the corresponding element in the
Fisher matrix). We have verified by numerical calculation that for a
Gaussian PDF, the cosmological dependence indeed improves constraints
on both $\Omega_m$ and $w$, in agreement with this argument.

\subsection{Why does non-Gaussianity of the PDF affect the constraints?}
\label{subsec:reason}

\begin{figure}
\begin{tabular}{c}
\rotatebox{-0}{\resizebox{80mm}{!}{\includegraphics{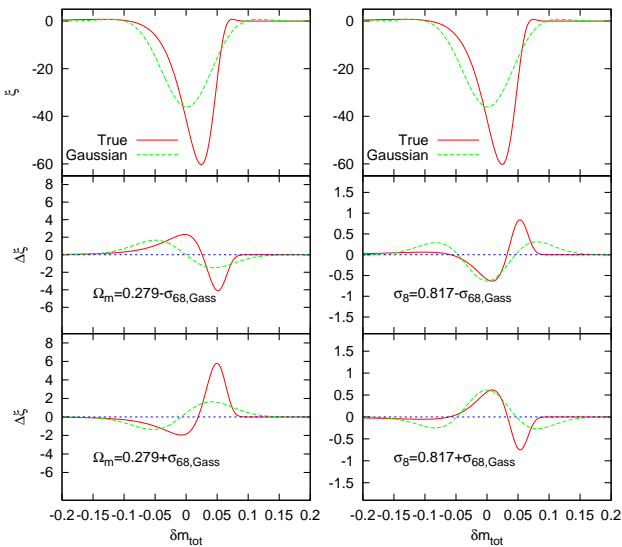}}}
\end{tabular}
\caption{The top panels show the contribution to $\xi$ (a
  $\chi^2$--like quantity; see text for the definition) at $z=1$ from
  events with different $\delta m_{\rm tot}$ (the deviation in the
  distance modulus from that in a homogeneous universe). The middle
  and bottom panels show the change in the $\xi$--contributions when
  the parameter is reduced or increased by one standard deviation
  $\sigma_{\rm 68,Gass}$. The left and right panels are for $\Omega_m$
  and $\sigma_8$, respectively.}
\label{fig:xiom}
\end{figure}

In this section, we discuss why the parameter constraints are affected
by the shape of the convergence PDF.  To illustrate this, we consider
a quantity $\xi$, defined as,
\begin{eqnarray}
\label{eqn:xi}
\xi\equiv \frac{d\langle\Xi\rangle}{d\delta m_{\rm tot}}= 
-2P_{\rm tot}(\delta m_{\rm tot}){\rm ln}P^{\prime}_{\rm tot}(\delta m^{\prime}_{\rm tot}),
\end{eqnarray}
where $\langle\Xi\rangle$ is the expectation value of $\Xi$ for one
event.  In equation (\ref{eqn:xi}), the factor $P_{\rm tot}(\delta
m_{\rm tot})$, computed with the true convergence PDF and with the
fiducial values of the cosmological parameters, is the probability
that an event has a (true) total deviation of $\delta m_{\rm
tot}$. The factor $-2{\rm ln}P^{\prime}_{\rm tot}(\delta
m^{\prime}_{\rm tot})$, computed with the convergence PDF adopted for
the likelihood analysis (i.e. either a Gaussian or the true PDF) and
with the varying cosmological parameters, represents the contribution
to $\Xi$ from a single event with this $\delta m_{\rm tot}$.  Note
that the residual distance modulus inferred in a test cosmology,
$\delta m^{\prime}_{\rm tot}$, differs from the true residual $\delta
m_{\rm tot}$, since the luminosity distances in the two cosmologies
are different. In the top left panel of Figure~\ref{fig:xiom}, we show
$\xi$ at $z=1$ as a function of the true residual $\delta m_{\rm
tot}$, where $\Omega_m$ is set at its fiducial value.

The change in $\xi$ when $\Omega_m$ is varied, $\Delta\xi$, is shown
in the left middle and left bottom panels, in which $\Omega_m$ is
decreased or increased by $\sigma_{\rm 68,Gass}$ ($=5.22 \times 10^
{-3}$, see Table~\ref{tbl:om1}), respectively. The main effect of
changing $\Omega_m$ is to shift the measured residual $\delta
m^{\prime}_{\rm tot}$ away from its true value of $\delta m_{\rm
tot}$, producing wave--like shapes in the curves $\Delta
\xi=\Delta\xi(\delta m_{\rm tot})$.  For example, assuming an
$\Omega_m$ larger than its fiducial value (bottom panel) increases the
apparent residuals $\delta m^{\prime}_{\rm tot}$ by reducing the
distances ($d_L(z)$) and therefore reducing $m_0$.  In other words,
the source is inferred to be too close, and the magnification is
inferred (statistically) to be smaller than the correct value.  This
corresponds to a reduced probability and increased $\xi$ at large
positive $\delta m_{\rm tot}$ where the $P_{\rm tot}(\delta m_{\rm
tot})$ decreases with $\delta m_{\rm tot}$ (i.e. if the source was
already strongly demagnified; see the dashed curve in Figure
~\ref{fig:pm}).  Conversely, it results in an enhanced probability and
a decreased $\xi$ at negative $\delta m_{\rm tot}$ (if the source was
strongly magnified, it is inferred to need less magnification, which
has a larger probability).  {\em The increase of $\xi$ at mildly
positive $\delta m_{\rm tot}$ and the decrease at mildly negative
$\delta m_{\rm tot}$ are both stronger in the case of true PDF.}  This
is because near $\delta m_{\rm tot}\approx 0$, the contribution to
$\xi$ is a much steeper function of $\delta m_{\rm tot}$ for the true
PDF, as seen clearly in the top panel.  Indeed, the change of $\Xi$,
i.e. the integral of $\Delta \xi$, is correspondingly also larger in
the case of the true PDF\footnote{Specifically, for the cases shown in
the left panels of Figure~\ref{fig:xiom}, we find $\Delta\Xi$= 0.0454
(true PDF, $\Omega_m$ reduced), 0.0232 (Gaussian PDF, $\Omega_m$
reduced),0.0506 (true PDF, $\Omega_m$ increased), 0.0214 (Gaussian
PDF, $\Omega_m$ increased).}, leading to an improved constraint on
$\Omega_m$.

The fact that the sharp features of the non-Gaussian lensing PDF near
its peak help tighten the sensitivity to $\Omega_m$ may be surprising,
since they are in contrast with the more widely discussed,
detrimental, effects of long non-Gaussian tails.  For example, the
mean of $N$ numbers, generated from a non--Gaussian distribution,
convergences less rapidly (than $1/\sqrt{N}$) to the correct value,
due to such shallow, non-Gaussian tails.  However, when one fits the
entire shape of the distribution (rather than utilizing just the
mean), then, intuitively, the benefits of sharp features are not
surprising.  For example, one can envision a hypothetical,
picket-fence-shaped distribution with the same r.m.s. as a smooth
Gaussian. It is easy to see that if the picket-fence pattern shifts
with a parameter, then this parameter can be arbitrarily tightly
constrained, in the limit that the picket-fence shapes become
arbitrarily sharp Dirac $\delta$--functions.

The right panels of Figure~\ref{fig:xiom} show the same quantities as
in the left panel, except for $\sigma_8$. Unlike $\Omega_m$,
$\sigma_8$ does not change the mean $\delta m^{\prime}_{\rm tot}$.  As
a result, in the Gaussian case, the curves of $\Delta \xi$ are
strictly symmetric about $\delta m_{\rm tot}=0$. The middle right and
the bottom right panels show that the non--Gaussianity again enhances
the sensitivity to $\sigma_8$ near the peak (for modest
demagnifications), whereas $\xi$ in the tail of the true PDF becomes
relatively insensitive to the change in $\sigma_8$.  Overall, this
loss of sensitivity in the tails is the larger effect, and it renders
$\Delta \Xi$ smaller than the Gaussian case\footnote{For the cases
shown in the right panels of Figure~\ref{fig:xiom}, $\Delta\Xi$=
0.0190 (true PDF, $\Omega_m$ reduced), 0.0289 (Gaussian PDF,
$\Omega_m$ reduced),0.0167 (true PDF, $\Omega_m$ increased), 0.0259
(Gaussian PDF, $\Omega_m$ increased).}. This explains why the
constraints on $\sigma_8$ becomes worse when the true PDF is used.

\subsection{How important is it to know the correct shape of the PDF?}
\label{subsec:bias}

\begin{table*}
  \caption{The peak of the maximum-likelihood estimator of $\Omega_m$
    and its 68\% and 95\% errors when the mock data, produced using
    the ``true'' PDF, is fitted by erroneously using either a Gaussian
    or a log-normal magnification PDF. The setup is otherwise the same
    as in Table~\ref{tbl:om1}.  }
    \label{tbl:wrongom}

\begin{center}
\begin{tabular}{c|c|c|c|c|c|c|c|c|c|c}
\hline \hline
 & \multicolumn{3}{c}{Gaussian
  PDF}& & &\multicolumn{3}{c}{Log-normal PDF}& & \\
$z_{\rm cent}$&$\nu_{\rm Gass}$&$\sigma_{68,\rm Gass}$&$\sigma_{95,\rm
 Gass}$& $\frac{\sigma_{68,\rm Gass}}{\sigma_{68,\rm true}}$&
 $\frac{\sigma_{95,\rm Gass}}{\sigma_{95,\rm true}}$&
$\nu_{\rm log}$&$\sigma_{68,\rm log}$ &$\sigma_{95,\rm
 log}$ & $\frac{\sigma_{68,\rm log}}{\sigma_{68,\rm true}}$&
 $\frac{\sigma_{95,\rm log}}{\sigma_{95,\rm true}}$ \\ 
\hline
1 & 0.277&$ 6.05 \times 10^ {-3}$&$ 1.32 \times 10^ {-2}$& 1.74& 1.86& 0.282&$ 3.73 \times 10^ {-3}$&$ 7.74 \times 10^ {-3}$& 1.07& 1.09\\
2 & 0.279&$ 7.83 \times 10^ {-3}$&$ 1.68 \times 10^ {-2}$& 1.25& 1.33& 0.279&$ 6.47 \times 10^ {-3}$&$ 1.35 \times 10^ {-2}$& 1.04& 1.07\\
3 & 0.279&$ 9.50 \times 10^ {-3}$&$ 1.94 \times 10^ {-2}$& 1.12& 1.11& 0.279&$ 8.78 \times 10^ {-3}$&$ 1.77 \times 10^ {-2}$& 1.03& 1.02\\
4 & 0.277&$ 1.22 \times 10^ {-2}$&$ 2.31 \times 10^ {-2}$& 1.07& 1.00& 0.274&$ 1.16 \times 10^ {-2}$&$ 2.34 \times 10^ {-2}$& 1.03& 1.01\\
\hline
\end{tabular}
\end{center}
\end{table*}

\begin{table*}
  \caption{Same as Table~\ref{tbl:wrongom}, except that $w$ is being constrained.   
}
    \label{tbl:wrongw}

\begin{center}
\begin{tabular}{c|c|c|c|c|c|c|c|c|c|c}
\hline \hline
 & \multicolumn{3}{c}{Gaussian
  PDF}& & &\multicolumn{3}{c}{Log-normal PDF}& & \\
$z_{\rm cent}$&$\nu_{\rm Gass}$&$\sigma_{68,\rm Gass}$&$\sigma_{95,\rm
 Gass}$& $\frac{\sigma_{68,\rm Gass}}{\sigma_{68,\rm true}}$&
 $\frac{\sigma_{95,\rm Gass}}{\sigma_{95,\rm true}}$&
$\nu_{\rm log}$&$\sigma_{68,\rm log}$ &$\sigma_{95,\rm
 log}$ & $\frac{\sigma_{68,\rm log}}{\sigma_{68,\rm true}}$&
 $\frac{\sigma_{95,\rm log}}{\sigma_{95,\rm true}}$ \\ 
\hline
1 & -1.003&$ 1.36 \times 10^ {-2}$&$ 2.62 \times 10^ {-2}$& 1.68& 1.57& -0.991&$ 8.15 \times 10^ {-3}$&$ 1.75 \times 10^ {-2}$& 1.01& 1.05\\
2 & -1.006&$ 2.90 \times 10^ {-2}$&$ 6.03 \times 10^ {-2}$& 1.25& 1.27& -1.000&$ 2.36 \times 10^ {-2}$&$ 4.94 \times 10^ {-2}$& 1.02& 1.05\\
3 & -1.012&$ 4.89 \times 10^ {-2}$&$ 9.71 \times 10^ {-2}$& 1.14& 1.10& -1.000&$ 4.37 \times 10^ {-2}$&$ 8.93 \times 10^ {-2}$& 1.02& 1.01\\
4 & -1.000&$ 7.23 \times 10^ {-2}$&$ 1.46 \times 10^ {-1}$& 1.06& 1.05& -0.988&$ 6.92 \times 10^ {-2}$&$ 1.39 \times 10^ {-1}$& 1.01& 1.00\\
\hline
\end{tabular}
\end{center}
\end{table*}

\begin{table*}
  \caption{Same as Table~\ref{tbl:wrongom}, except that $\sigma_8$ is constrained and there are 400 events in each realization.  
}
    \label{tbl:wrongs8}

\begin{center}
\begin{tabular}{c|c|c|c|c|c|c|c|c|c|c}
\hline \hline
 & \multicolumn{3}{c}{Gaussian
  PDF}& & &\multicolumn{3}{c}{Log-normal PDF}& & \\
$z_{\rm cent}$&$\nu_{\rm Gass}$&$\sigma_{68,\rm Gass}$&$\sigma_{95,\rm
 Gass}$& $\frac{\sigma_{68,\rm Gass}}{\sigma_{68,\rm true}}$&
 $\frac{\sigma_{95,\rm Gass}}{\sigma_{95,\rm true}}$&
$\nu_{\rm log}$&$\sigma_{68,\rm log}$ &$\sigma_{95,\rm
 log}$ & $\frac{\sigma_{68,\rm log}}{\sigma_{68,\rm true}}$&
 $\frac{\sigma_{95,\rm log}}{\sigma_{95,\rm true}}$ \\ 
\hline
1 & 0.790&$ 5.02 \times 10^ {-2}$&$ 1.06 \times 10^ {-1}$& 1.67& 1.78& 0.742&$ 2.53 \times 10^ {-2}$&$ 4.93 \times 10^ {-2}$& 0.84& 0.83\\
2 & 0.814&$ 3.68 \times 10^ {-2}$&$ 7.02 \times 10^ {-2}$& 1.32& 1.32& 0.790&$ 2.66 \times 10^ {-2}$&$ 5.04 \times 10^ {-2}$& 0.95& 0.95\\
3 & 0.814&$ 3.25 \times 10^ {-2}$&$ 6.47 \times 10^ {-2}$& 1.15& 1.13& 0.802&$ 2.76 \times 10^ {-2}$&$ 5.56 \times 10^ {-2}$& 0.97& 0.97\\
4 & 0.826&$ 4.13 \times 10^ {-2}$&$ 8.06 \times 10^ {-2}$& 1.09& 1.08& 0.814&$ 3.80 \times 10^ {-2}$&$ 7.47 \times 10^ {-2}$& 1.00& 1.00\\
\hline
\end{tabular}
\end{center}
\end{table*}

Our results above show that the shape of the lensing PDF has a
non-negligible effect on the parameter constraints.  However, the
analysis above has assumed that the shape of the lensing PDF is
precisely known.

A related question of interest, therefore, is: how accurately do we
need to know the lensing PDF, in order to avoid either a bias in the
inferred parameters, or a mis-estimation of its confidence levels?

In order to investigate this issue, we here create mock data-sets with
the ``true'' lensing PDF from DO06, but we fit them using ``wrong''
PDFs. The two ``wrong'' PDFs we chose are a Gaussian (representing an
extremely wrong case), and a log-normal distribution (which represents
a more reasonable estimate of the current uncertainty about the
convergence PDF). The results for $\Omega_m$, $w$ and $\sigma_8$ are
shown in Tables~\ref{tbl:wrongom}--\ref{tbl:wrongs8}.

Tables~\ref{tbl:wrongom} shows that the maximum--likelihood estimate
(MLE) of $\Omega_m$ remains essentially unbiased: the MLE distribution
peaks at values that are offset from the true value of
$\Omega_m=0.279$ by at most $\approx 1/3$rd of the 68\% confidence
level.  This result is in agreement with DV06, and is not surprising,
since the peak of the best--fit $\Omega_m$ distribution is determined
primarily by the luminosity distance (or the homogeneous distance
modulus $m_0$), rather than the shape of the $\delta m$--distribution.
The variance of the best--fit $\Omega_m$, however, becomes larger when
the Gaussian or log-normal PDF is used in the likelihood analysis. The
difference in the lowest redshift bin is $\sim 80\%$ for a Gaussian
PDF, but only $\sim 10\%$ for the log--normal PDF.  This question has
also been investigated in the context of SNe Ia by DV06, who found
that the variance is almost unchanged.  We have confirmed their
conclusion by running a similar Monte-Carlo simulation for a mock SNe
Ia survey.  The reason why the constraints on $\Omega_m$ from standard
sirens, in contrast to standard candles, are obviously affected by the
choice of the convergence PDF, is the large skewness (seen in
Fig.~\ref{fig:pm}) in the error distribution of standard sirens (see
more explicit demonstration of this point below).  From
Table~\ref{tbl:wrongom}, it is also clear that the improvement of the
constraints on $\Omega_m$ decreases with increasing source
redshift. This feature could also be explained by the decrease in the
skewness of the distribution with increasing source redshift. Here,
the change is not driven by the intrinsic dispersion, but by the
lensing dispersion. As we show in Figure~\ref{fig:pk}, the convergence
PDF is more skewed at low redshift, making the effect of
non-Gaussianity more prominent.

The conclusions for $w$ are qualitatively very similar, except that
the differences caused by the ``wrong'' PDFs are smaller, as shown in
Table~\ref{tbl:wrongw}. In fact, the constraints with the log-normal
PDF are almost identical to those from the ``true'' PDF, with a
negligible bias, and with an overestimate of the 68\% error by only
2\%.  We expect these qualitative conclusions, from the examples of
$\Omega_m$ and $w$, to remain valid for other parameters such as the
Hubble constant $h$, or the dark energy equation of state evolution
parameter $dw/dz$, whose impact is primarily through $d_L(z)$, rather
than the shape of the lensing PDF.

\begin{figure}
\begin{tabular}{c}
\rotatebox{-0}{\resizebox{90mm}{!}{\includegraphics{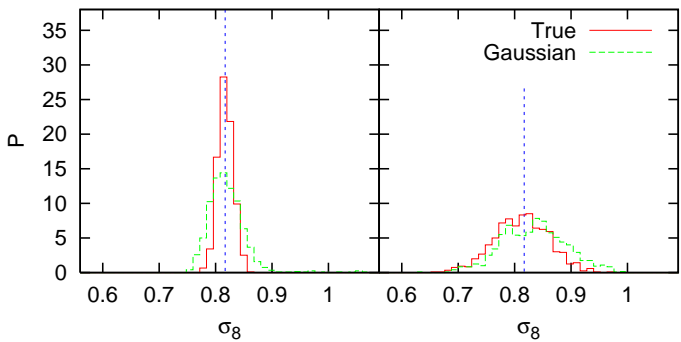}}}
\end{tabular}
\caption{Distributions of the best--fit $\sigma_8$ values, obtained by
  fitting the ``true'' magnification PDF with erroneous Gaussian PDFs,
  in two Monte-Carlo simulations that only differ in the size of the
  intrinsic scatter. The left (right) panel shows the results when the
  intrinsic scatter is typical of standard sirens (standard
  candles). Two thousand events are generated in each
  realization. Note that $\sigma_8$ is negatively biased for standard
  sirens, while it is positively biased for SNe (see text and
  Figure~\ref{fig:xis8} for an explanation of this result).}
\label{fig:bias}
\end{figure}

Interestingly, as shown in Table~\ref{tbl:wrongs8}, for $\sigma_8$,
the erroneous use of the Gaussian PDF also degrades the constraints,
while the log-normal PDF gives rise to better constraints, but causes
a clear negative bias. At the lowest redshift, the bias is $\sim
0.07$, exceeding the forecast 68\% constraint.  The still larger, and
positive, bias of $\sigma_8$ for the Gaussian PDF seen in DV06 was not
found in our calculations; instead, the peak is slightly negatively
biased in the lowest redshift bin (we find the mean, however, is
almost unbiased).

Investigations of the seeming inconsistency with DV06 revealed,
initially to our own surprise, that the sign of the $\sigma_8$--bias
is affected by the size of the intrinsic scatter, in addition to the
skewness of the lensing PDF. To illustrate this point, in
Figure~\ref{fig:bias} we plot the distributions of the best--fit
$\sigma_8$ from two Monte-Carlo simulations that only differ in the
amount of the intrinsic scatter -- the left panel assumes the
intrinsic dispersion of a typical standard siren, while the right
panel assumes that of a typical standard candle ($\sigma_{\rm
int}=0.1$, as before).  As the right panel of this figure shows, in
addition to increasing the variance of the inferred $\sigma_8$
distribution, a larger intrinsic dispersion also shifts the peak
position to a higher value, in the case of the Gaussian PDF. This
positive bias in the SN sample has a value of $\Delta\sigma_8\approx
0.017$.

\begin{figure}
\begin{tabular}{c}
\rotatebox{-0}{\resizebox{80mm}{!}{\includegraphics{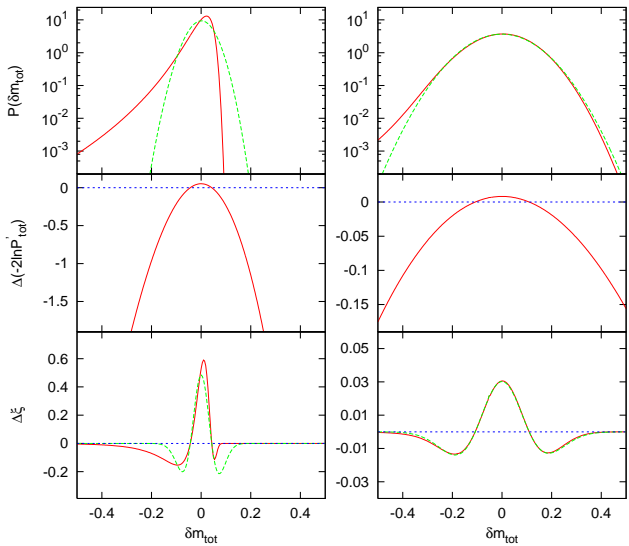}}}
\end{tabular}
\caption{The top panels show the probability distributions of $\delta
  m_{\rm tot}$ at $z=1$. The solid and dashed curves assume that the
  true convergence PDF follows $P_{\rm Do}$ or a Gaussian,
  respectively. The middle panels show the change in $-2{\rm
  ln}P^{\prime}_{\rm tot}(\delta m^{\prime}_{\rm tot})$ when
  $\sigma_8$ is increased from the fiducial value of 0.817 to 0.834
  (the mean of $\sigma_8$ from SNe Ia simulations when Gaussian PDF is
  used in the likelihood analysis). The bottom panels show the
  corresponding change in the $\chi^2$--like quantity $\xi$. The left
  and right panels are for standard sirens and SNe Ia, respectively.}
\label{fig:xis8}
\end{figure}

To understand why the $\sigma_8$--bias changes sign as the intrinsic
scatter is increased, we again consider the quantity $\xi$ and its
dependence on $\sigma_8$.  In the bottom panels of
Figure~\ref{fig:xis8}, we show $\Delta \xi$ ($\equiv
\xi(\sigma_8=0.834)-\xi({\rm fiducial}~ \sigma_8=0.817)$, solid
curves) for the case of the erroneous Gaussian PDF; the left and right
panels are again for standard sirens and for SNe Ia, respectively.  In
agreement with the results of the Monte-Carlo simulations, $\Delta
\Xi$ from the integration of $\Delta \xi$ is negative for SNe Ia,
meaning the bias is positive, while $\Delta \Xi$ for standard sirens
is positive, consistent with no bias.  The top and bottom panels show
$P_{\rm tot}(\delta m_{\rm tot})$ and $\Delta(-2{\rm
ln}P^{\prime}_{\rm tot}(\delta m^{\prime}_{\rm tot}))$, the two
multiplicative factors comprising $\Delta \xi$.  For comparison, we
also show, by the dashed curves, the same quantities in a ``true
Gaussian'' case, where the ``true'' convergence PDF (in addition to
the PDF employed in the likelihood analysis) is assumed to be
Gaussian, and therefore the estimate of $\sigma_8$ should be unbiased.

Compared with the Gaussian ``true'' distribution, the skewed ``true''
distribution produces a much higher probability of strongly magnified
events (solid curves in the top panels). These events contribute
negative $\Delta \xi$, i.e., they favor larger $\sigma_8$ values (the
solid curve is largely below the dashed curve at $\delta m_{\rm tot} <
0$).  This is the effect that dominates in the case of SNe, as already
argued by DV06.  However, when the intrinsic dispersion is small, this
negative contribution to $\Xi$ is, in fact, balanced by the decrease
of the probability at $\delta m_{\rm tot} \gsim 0$.  In other words,
the large $\sigma_8$ value that is favored by the events in the
strongly magnified tail, is even more strongly disfavored by the lack
of corresponding mildly demagnified events -- leading to an overall
{\em negative} bias in $\sigma_8$. In the SN case, when the intrinsic
dispersion is large, there is no longer a lack of mildly demagnified
events to be disfavored, because the expected number of these events
is still dominated by the intrinsic scatter, and is insensitive to
$\sigma_8$ --- the net effect, in this case, is the positive bias in
$\sigma_8$, driven by the highly magnified events.

\subsection{Varying two cosmological parameters}
\label{subsec:2p}

\begin{figure}
\begin{tabular}{c}
\rotatebox{-0}{\resizebox{150mm}{!}{\includegraphics{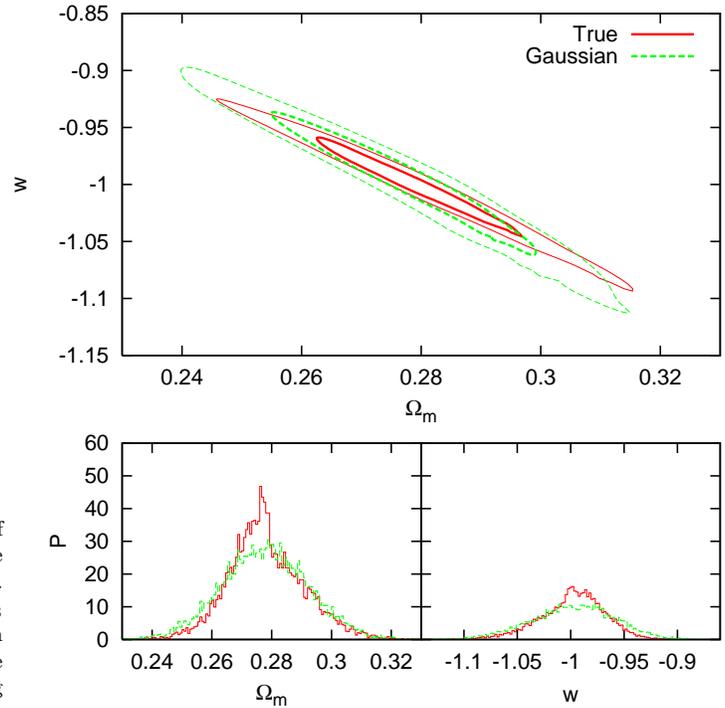}}}
\end{tabular}
\caption{{\it Upper panel}: Contours enclosing 68\% and 95\% of the
  best--fit values on $\Omega_m$ and $w$ in the lowest redshift bin
  for the true (solid line) and Gaussian (dashed line) PDF.  In each
  case, $10^4$ realizations were produced of mock samples containing
  100 standard sirens, and $\Omega_m$ and $w$ were fit simultaneously,
  with all other parameters fixed at their fiducial values. {\it
  Bottom left panel}: Distribution of best--fit $\Omega_m$,
  marginalized over $w$. {\it Bottom right panel}: Distribution of
  best--fit $w$, marginalized over $\Omega_m$. }
\label{fig:omw}
\end{figure}

\begin{figure}
\begin{tabular}{c}
\rotatebox{-0}{\resizebox{150mm}{!}{\includegraphics{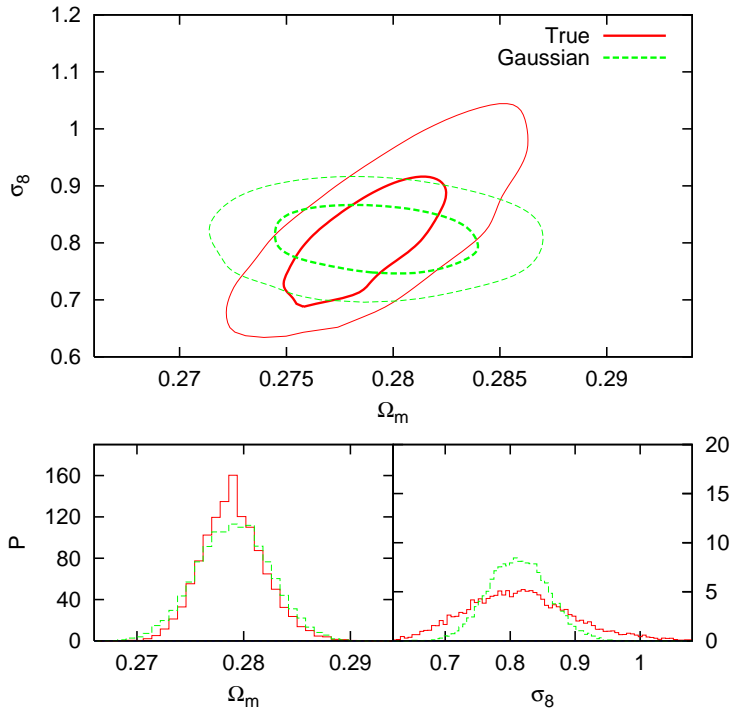}}}
\end{tabular}
\caption{Same as Figure~\ref{fig:omw}, except that $\Omega_m$ and
  $\sigma_8$ are being simultaneously constrained.}
\label{fig:oms8}
\end{figure}

\begin{table*}
  \caption{The peak, 68\% and 95\% marginalized errors of $\Omega_m$
    and $w$ for Gaussian and true PDFs. Each realization produces 100
    events.} 
  \label{tbl:omw}
\begin{center}
\begin{tabular}{c|c|c|c|c|c|c|c|c}
\hline \hline
 & \multicolumn{3}{c}{Gaussian
  PDF}&\multicolumn{3}{c}{True PDF}& & \\
Parameter&$\nu_{\rm Gass}$&$\sigma_{68,\rm Gass}$&$\sigma_{95,\rm
 Gass}$&$\nu_{\rm true}$&$\sigma_{68,\rm true}$ &$\sigma_{95,\rm
 true}$ & $\frac{\sigma_{68,\rm Gass}}{\sigma_{68,\rm true}}$&
 $\frac{\sigma_{95,\rm Gass}}{\sigma_{95,\rm true}}$\\ 
\hline
$\Omega_m$ & 0.280&$ 1.44 \times 10^ {-2}$&$ 2.87 \times 10^ {-2}$& 0.277&$ 1.22 \times 10^ {-2}$&$ 2.69 \times 10^ {-2}$& 1.18& 1.07\\
$w$ & -0.986&$ 3.69 \times 10^ {-2}$&$ 7.78 \times 10^ {-2}$& -1.000&$ 2.92 \times 10^ {-2}$&$ 6.18 \times 10^ {-2}$& 1.26& 1.26\\
\hline
\end{tabular}
\end{center}
\end{table*}

\begin{table*}
  \caption{The peak, 68\% and 95\% marginalized errors of $\Omega_m$
    and $\sigma_8$ for Gaussian and true PDFs. Each realization produces 100
    events.} 
  \label{tbl:oms8}
\begin{center}
\begin{tabular}{c|c|c|c|c|c|c|c|c}
\hline \hline
 & \multicolumn{3}{c}{Gaussian
  PDF}&\multicolumn{3}{c}{True PDF}& & \\
Parameter&$\nu_{\rm Gass}$&$\sigma_{68,\rm Gass}$&$\sigma_{95,\rm
 Gass}$&$\nu_{\rm true}$&$\sigma_{68,\rm true}$ &$\sigma_{95,\rm
 true}$ & $\frac{\sigma_{68,\rm Gass}}{\sigma_{68,\rm true}}$&
 $\frac{\sigma_{95,\rm Gass}}{\sigma_{95,\rm true}}$\\ 
\hline
$\Omega_m$ & 0.280&$ 3.46 \times 10^ {-3}$&$ 6.92 \times 10^ {-3}$& 0.279&$ 2.92 \times 10^ {-3}$&$ 6.03 \times 10^ {-3}$& 1.19& 1.15\\
$\sigma_8$ & 0.814&$ 4.64 \times 10^ {-2}$&$ 9.25 \times 10^ {-2}$& 0.820&$ 7.99 \times 10^ {-2}$&$ 1.65 \times 10^ {-1}$& 0.58& 0.56\\
\hline
\end{tabular}
\end{center}
\end{table*}

In our calculations above, we varied a single parameter, while all
others were held fixed.  To take into account the correlations among
different cosmological parameters, we next consider simultaneous
constraints on two cosmological parameters. Ideally, all of the
cosmological parameters should be allowed to vary simultaneously, for
a full study of correlations.  Unfortunately, with our simple
Monte-Carlo method, this is computationally prohibitively expensive
(and would require the use of Monte Carlo Markov Chains, or a similar
technique).  In this study, for simplicity, we instead chose to study
only two typical correlations as illustrative examples. One of the
correlations, between $\Omega_m$ and $w$, is expected to be large,
since as we have shown, both parameters draw their constraints from
the mean distance modulus. The other correlation, between $\Omega_m$
and $\sigma_8$, is expected to be small, as the constraints on
$\sigma_8$ are purely from the cosmological dependence of the shape of
the convergence PDF. We focus on the lowest redshift bin of $z_{\rm
cent}=1$, where the assumed shape of the convergence PDF makes the
largest difference. One hundred events are generated in each
realization.

The joint 2-D confidence contours and the marginalized 1-parameter
distributions are plotted in Figure~\ref{fig:omw} (for simultaneous
constraints on $\Omega_m$ and $w$) and in Figure~\ref{fig:oms8} (for
simultaneous constraints on $\Omega_m$ and $\sigma_8$). The values of
the marginalized errors are also listed in Tables~\ref{tbl:omw} and
\ref{tbl:oms8}.

Figure~\ref{fig:omw} confirms that $\Omega_m$ and $w$ are strongly
anti--correlated. As in the single parameter case, $\Omega_m$ and $w$
are both more tightly constrained with the true PDF, by 18\% for
$\Omega_m$ and 26\% for $w$ (evaluated using their marginalized 68\%
errors). The area enclosed by the 68\% (95\%) contour is larger in the
Gaussian PDF case by 117\% (92\%).  The raw sensitivity to the
combination of $\Omega_m$ and $w$ is therefore improved, by the
presence of non-Gaussianity, by more than a factor of two.

Figure~\ref{fig:oms8} shows that the correlation between $\Omega_m$
and $\sigma_8$ depends on the shape of the convergence PDF. With a
Gaussian PDF, the parameters are almost uncorrelated\footnote{In fact,
$\Omega_m$ and $\sigma_8$ are slightly anti--correlated, since both
parameters affect the lensing dispersion.}, while with the true PDF,
they are positively correlated. Here we offer a heuristic explanation
for this positive correlation: with larger $\sigma_8$, the convergence
dispersion is increased, making the convergence PDF more skewed, and
its peak shift to lower $\kappa$; to fit the same $m_{\rm tot}$, this
requires a larger $\Omega_m$ to compensate. The marginalized error on
$\Omega_m$ is still smaller with the true PDF, but the difference of
the 68\% errors, only 19\%, is considerably smaller than in the single
parameter case. At the same time, the constraints on $\sigma_8$ are
slightly more degraded than those with a Gaussian PDF. Here, the area
enclosed by the 68\% (95\%) contour is smaller in the Gaussian PDF
case by 6\% (10\%).

\subsection{The importance of the number of events}
\label{subsec:Nevents}

Another interesting question to ask is: how do our results -- in
particular, the tightening or degradation in the constraints caused by
non--Gaussianities -- depend on the number of standard siren events
$N_{\rm event}$?  Depending on what statistic is being used,
non-Gaussianities can become increasingly unimportant as $N_{\rm
event}\rightarrow\infty$. This is the case, for example, when one
considers the distribution of the {\em mean} magnification
$\langle\delta m\rangle$ from $N_{\rm event}$ events (as is well known
from the central limit theorem; see, e.g., Figure 1 in Holz \& Linder
2005).  This statistic, however would be relevant only if one simply
ignored lensing, and performed a least-squares fit to the observed
Hubble diagram.  This is different from the maximum-likelihood
estimator we adopted here, which includes information beyond the mean
magnification, and utilizes the full shape of the PDF.  In fact, we
expect that the MLE has a distribution for which the non-Gaussianity
does {\em not} average away in the limit of a large number of events.

In Table~\ref{tbl:nevent}, we show the ratios of the 68\% errors in
the true and the Gaussian cases, when the number of events in each
realization is increased and decreased, by a factor of 4, from the
values adopted in \S~\ref{subsec:1p}. As the table shows, the ratios
are consistent with no dependence on $\sqrt{N_{\rm event}}$ (while the
absolute errors, which are not shown in the table, were found to scale
approximately as $\sqrt{N_{\rm event}}$). Therefore, the general
conclusion of this study, that the non-Gaussian shape non-trivially
affects cosmological constraints, is valid independent of $N_{\rm
event}$. This also demonstrates the importance of choosing an optimal
statistic: the information from the skewness in the PDF would be lost
if only the mean distance modulus were used.

\begin{table}
  \caption{The ratios of the 68\% errors on the individual
 cosmological parameters in the true and in the Gaussian cases, when
 the number of events in each realization is increased or decreased by
 a factor of 4 relative to our fiducial choices.}
  \label{tbl:nevent}
\begin{center}
\begin{tabular}{c|c|c|c}
\hline \hline
$N_{\rm event}$&$\times 0.25$&$\times 1$&$\times 4$\\
\hline
$\Omega_m$& 1.57& 1.50& 1.55\\
$w$& 1.72& 1.78& 1.70\\
$\sigma_8$& 0.76& 0.83& 0.82\\
\hline
\end{tabular}
\end{center}
\end{table}

\subsection{The expected constraints from a {\it LISA} standard siren sample}
\label{subsec:LISAerrors}

\begin{figure}
\begin{tabular}{c}
\rotatebox{0}{\resizebox{80mm}{!}{\includegraphics{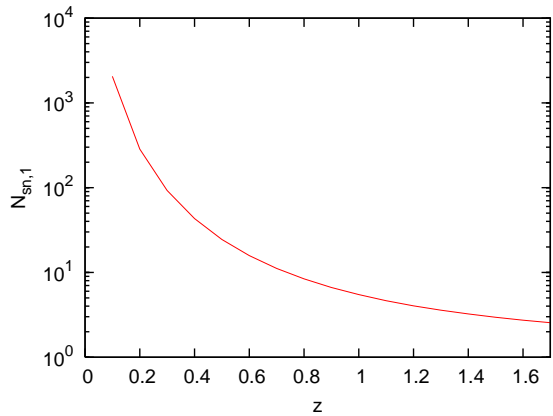}}}
\end{tabular}
\caption{A crude estimate for the number of SNe whose statistical
  constraining power is equivalent to that of a single standard siren.
  We have assumed that the magnitude errors of the two types of
  sources follow the same distribution, with the amplitude given by
  the convolution of intrinsic and lensing errors.}
\label{fig:nevent}
\end{figure}

What is the actual expected error from {\it LISA} standard sirens on
cosmological parameters, in the presence of lensing?  Since we have
not obtained these errors in the full multi--dimensional parameter
space, we cannot fully answer this question.  However, the same
question has been studied exhaustively in the context of SNe -- and,
since the SNe constraints are from the same observable (the Hubble
diagram), it makes sense to quote the expectations from standard
sirens relative to those from a SN sample.  If the magnitude errors
had the same distribution for these two types of sources, with
standard deviations of $\sigma_{\rm ss}$ and $\sigma_{\rm sn}$, then
the statistical constraints from a single standard siren would be
equivalent to that from $N_{\rm sn,1}=(\sigma_{\rm ss}/\sigma_{\rm
sn})^{-2}$ SNe.  Assuming a constant $\sigma_{\rm sn}\approx 0.1$ mag,
and also that the lensing and intrinsic errors add in quadrature, in
Figure~\ref{fig:nevent} we show $N_{\rm sn,1}$.  As the intrinsic
errors of standard sirens approach those of SNe, their relative
advantage decreases monotonically with redshift, with $N_{\rm
sn,1}\approx 93 \rightarrow 5.5\rightarrow 2.5$ at $z\approx 0.3
\rightarrow 1 \rightarrow 1.7$.  (Peculiar velocities will limit the
constraints from standard sirens below redshifts $z\lsim 0.3$.)  The
main result of this paper could be stated by saying that at $z=1$,
these numbers for $N_{\rm sn}$ are increased by a factor of
$1.5^2=2.25$ for $\Omega_m$, and by a factor of $1.8^2=3.25$ for $w$;
whereas they are decreased by a factor of $0.8^2=0.64$ for $\sigma_8$.
Note that while the actual numbers for $N_{\rm sn,1}$ are only crude
estimates, these increases/decreases, which are due to the
non--Gaussian shape of the lensing PDF, should be robust. At higher
redshifts, these increases/decreases disappear.

These estimates apply to a single redshift, and to a single
cosmological parameter.  The constraints from an actual set of {\em
LISA} standard sirens will, of course, depend on the redshift
distribution of the sources.  If most of the standard sirens are at
high redshifts ($z\gsim 3$; e.g., in Sesana 2004) then our results
suggest that there will be no significant overall improvements from
non-Gaussianity for the set of events.\footnote{Of course, the fact
that the standard siren sample extends to high redshift is still a
useful complement to SN datasets, which are not expected to reach
beyond $z\sim 1.7$ (Aldering et al. 2004).}  If there are more events
at low redshift, then non-Gaussianity will improve the constraints
significantly. Having low-$z$ events remains a possibility (see, e.g.,
discussion of some of the uncertainties, and related references, in
Lippai et al. 2009). Holley-Bockelmann et al. (2010) recently found,
using cosmological simulations of the BH merger trees, that most
LISA-detectable merger events are below $z=2$ (albeit with high mass
ratios, so their intrinsic $d_L$ error will be worse than for the
equal--mass binaries adopted here).  Finally, even if most events are
at $z\gsim 4$, non--Gaussianities can still improve the overall
constraints when the {\em LISA} data is combined with information from
still higher redshift, such as cosmic microwave background (CMB)
anisotropies.  Since the CMB fixes the luminosity distance at
$z\approx 1,000$, the lowest--redshift standard sirens will have the
longest ``lever--arm'' and the contribution from these events to the
combined {\em LISA}+CMB constraints will be enhanced significantly (a
point emphasized in a different context by Hu \& Haiman 2003).

\section{Conclusions}
\label{sec:conclusion}

In this paper, we have shown that the sensitivity of low--redshift
($z\sim 1$) standard sirens to the parameters $\Omega_m$ and $w$ are
tightened, by a factor of $1.5-1.8$, by the non--Gaussian shape of the
lensing magnification PDF, relative to a Gaussian PDF with the same
variance.  When these two parameters are constrained simultaneously,
the improvement of the constraints, attributable to the skewness
alone, is further enhanced, owing to the correlation between the
parameters.  We expect similar conclusions to hold for other
parameters whose constraints are driven primarily by the changes they
induce in the luminosity distance $d_L(z)$.  Interestingly, the
constraint on $\sigma_8$, which comes from the shape of the PDF
itself, is, however, degraded by a factor of $\approx 0.8$.  In our
study, we relied on the shape of the lensing PDF described by the
fitting formulae in Das \& Ostriker (2006). This shape is somewhat
less skewed than the simulation results, suggesting that the
improvement / degradation of the constraints has likely been
underestimated.

The improvements for $\Omega_m$ and $w$ are comparable to those that
may be available from correlations with lensing measurements on larger
scales, or from directly subtracting the lensing contribution of
foreground galaxies. However, these corrections from non-Gaussianities
are ``automatically'' available, as long as the lensing PDF can be
modeled ab--initio.  We also found that at higher redshift, the
effects from non--Gaussianity are less pronounced due to the reduced
skewness of the lensing PDF.  Overall, our results highlight the
importance of obtaining an accurate and reliable PDF of the
point--source lensing magnification, in order to realize the full
potential of standard sirens as cosmological probes.

\vspace{-0.5\baselineskip}

\section*{Acknowledgments}

This work was supported by the Pol\'anyi Program of the Hungarian
National Office for Research and Technology (NKTH) and by the NASA
ATFP grant NNXO8AH35G.


\label{lastpage}
\end{document}